# Low Dimensional Assemblies of Magnetic MnFe$_2$O$_4$ Nanoparticles and Direct *In Vitro* Measurements of Enhanced Heating Driven by Dipolar Interactions: Implications for Magnetic Hyperthermia

*Beatriz Sanz[†,§,◊,‡], Rafael Cabreira-Gomes[†,℘,‡], Teobaldo E. Torres[†,¶,⁎], Daniela P. Valdés[⁎], Enio Lima Jr.[⁎], Emilio De Biasi[⁎], Roberto D. Zysler[⁎], M. Ricardo Ibarra[†,§] and Gerardo F. Goya[†,§,*].*

[†] Instituto de Nanociencia de Aragón, Universidad de Zaragoza, Zaragoza, Spain

[§] Condensed Matter Physics Department, University of Zaragoza, Zaragoza, Spain

[¶] Laboratorio de Microscopias Avanzadas, Universidad de Zaragoza, Zaragoza, Spain

[⁎] Comisión Nacional de Investigaciones Científicas y Técnicas (CONICET), Centro Atómico Bariloche, Bariloche, Argentina.

ABSTRACT Magnetic fluid hyperthermia (MFH), the procedure of raising the temperature of tumor cells using magnetic nanoparticles (MNPs) as heating agents, has proven successful in treating some types of cancer. However, the low heating power generated under physiological conditions makes necessary a high local concentration of MNPs at tumor sites. Here, we report how the *in vitro* heating power of magnetically soft $MnFe_2O_4$ nanoparticles can be enhanced by intracellular low-dimensional clusters through a strategy that includes: a) the design of the MNPs to retain Néel magnetic relaxation in high viscosity media, and b) culturing MNP-loaded cells under magnetic fields to produce elongated intracellular agglomerates. Our direct *in vitro* measurements demonstrated that the specific loss power (SLP) of elongated agglomerates (SLP=576±33 W/g) induced by culturing BV2 cells *in situ* under a dc magnetic field was increased by a factor of 2 compared to the SLP=305±25 W/g measured in aggregates freely formed within cells. A numerical mean-field model that included dipolar interactions quantitatively reproduced the SLPs of these clusters both in phantoms and *in vitro*, suggesting that it captures the relevant mechanisms behind power losses under high-viscosity conditions. These results indicate that *in situ* assembling of MNPs into low-dimensional structures is a sound possible way to improve the heating performance in MFH.

## 1. Introduction

The number of procedures that use magnetic nanoparticles (MNPs) in biomedical and clinical protocols for diagnosis and therapy continues to grow steadily. One of these enthralling applications, known as magnetic fluid hyperthermia (MFH), is an oncological therapy aimed at delivering heat from locally injected MNPs to cancer cells.[1] The MNPs act as nanoheaters that absorb energy from a remotely applied radiofrequency magnetic field (frequency $f \approx$ 100-200 kHz) and release heat that locally increases the temperature of cancer cells to 42-46 ºC. Although MFH has already been applied in clinical trials [2,3] and has proven successful in treating some types of cancer [4], a main drawback of this therapy is the low heating power of MNPs under physiological conditions and clinically acceptable magnetic fields. Ultimately, despite the quite high specific loss power (SLP) values of up to $10^4$ W/g that can be obtained in *as-prepared* colloidal MNPs,[5] these values drop to < 200 W/g under *in vitro* or *in vivo* conditions [6]. The improved heating performance of bacterial magnetosomes with chain-like organization has been reported [7] in connection with their particle morphology, although little attention was paid to the role of the local dipolar magnetic fields that are known to be important for these 1D structures[8]. It has become clear in recent years that power absorption is hindered under physiological conditions through a) the spontaneous aggregation of the MNPs in protein-rich media and b) the high viscosity of the intracellular medium and/or attachment of MNPs to membranes[6, 9]. Each of these effects prevents the physical movement of the particles (Brownian relaxation) and lowers the heating efficiency of the MNPs. The agglomeration of MNPs that naturally occurs in many biological environments also brings magnetic dipolar interactions between MNPs into play. While the stalling of Brown relaxation and consequent decrease in SLP due to high-viscosity environments [10,11] has been well-documented in the literature, there is still controversy regarding the effects of dipolar interactions on SLP when intracellular agglomeration occurs.[12,13,14,15,16,17] Recent theoretical works on cluster geometry reflect this lack of consensus: while some authors have found that linear structures can increase the SLP of MNPs,[18,19] some others have reported opposite results, i.e., that the formation of linear chain aggregates significantly decreases the heating efficiency.[20,21,22] This state of debate is partially related to the scarce amount of direct experimental SLP data *in vitro* or *in vivo*.[23,24] The need for such measurements is indisputable since only from these data will we be able to calculate *a priori* the suitable clinical doses required for any particular cancer model to be treated[25].

The experiments and numerical models reported in this work address the above issues and establish that it is possible to enhance the heating efficiency *in vitro* by culturing MNP-loaded cells under an external constant magnetic field $H_{DC}$, which induces elongation of the intracellular agglomerates naturally occurring in cells. We performed systematic measurements on these elongated agglomerates of MNPs in tissue-equivalent gel phantoms and compared the results to similar *in situ*-formed agglomerates within the cytoplasm of BV2 microglial cells cultured under an applied dc field. We found clear evidence that intracellular clusters grown *in vitro* under a constant applied field $H_{DC}$ have an elongated shape with uniaxial symmetry, in contrast with the spherically shaped agglomerates that form spontaneously (i.e., with $H_{DC} = 0$) during uptake. The agglomerates formed under $H_{DC}$ fields showed increased SLP values in synthetic gelatin phantoms as well as *in vitro* experiments. Using a simple model of dipolar interactions on agglomerates with cylindrical symmetry, we performed numerical calculations to quantitatively reproduce the SLP of both the phantoms and *in vitro* systems, which confirmed the observed increase in SLP when elongated agglomerates were formed. This model showed that the increase in SLP originates from the increase of the magnetic anisotropy energy barrier to the effective magnetic anisotropy of the MNPs driven by the local dipolar field within the linear aggregates, which causes the increase of the hysteresis area during ac cycles and consequently increases the measured SLP.

## 2. Materials and Methods

The nanoparticles used in this work were synthesized through the co-precipitation method developed by Massart and Cabuil [26] based on the precipitation of an iron(II) salt ($FeSO_4$) in the presence of a base (NaOH) and a mild oxidant ($KNO_3$) and adapted accordingly to produce $MnFe_2O_4$ (manganese ferrite) MNPs.[27,28,29] First, 135 mL of NaOH buffer (0.8 M) was heated at 95 °C under $N_2$ flow in a refluxed three-necked quartz flask. In a second step the precursors

FeSO$_4$·4H$_2$O (0.2 M) and MnCl$_2$·6H$_2$O (0.1 M) were added to a flask with 15 mL of milli-Q water, 0.125 g of PEI (branched, 25 kDa) and 6 μl of HCl (37%), followed by 5-10 min of sonication. The resulting solution was swiftly added to the buffer in the quartz flask at 95 ºC and kept at that temperature under mechanical stirring (200 rpm) for 30 min. After that time, the particles were washed 2-3 times with deionized water until pH = 6.8-7.2 was reached. This procedure is schematized in **Figure *1***.

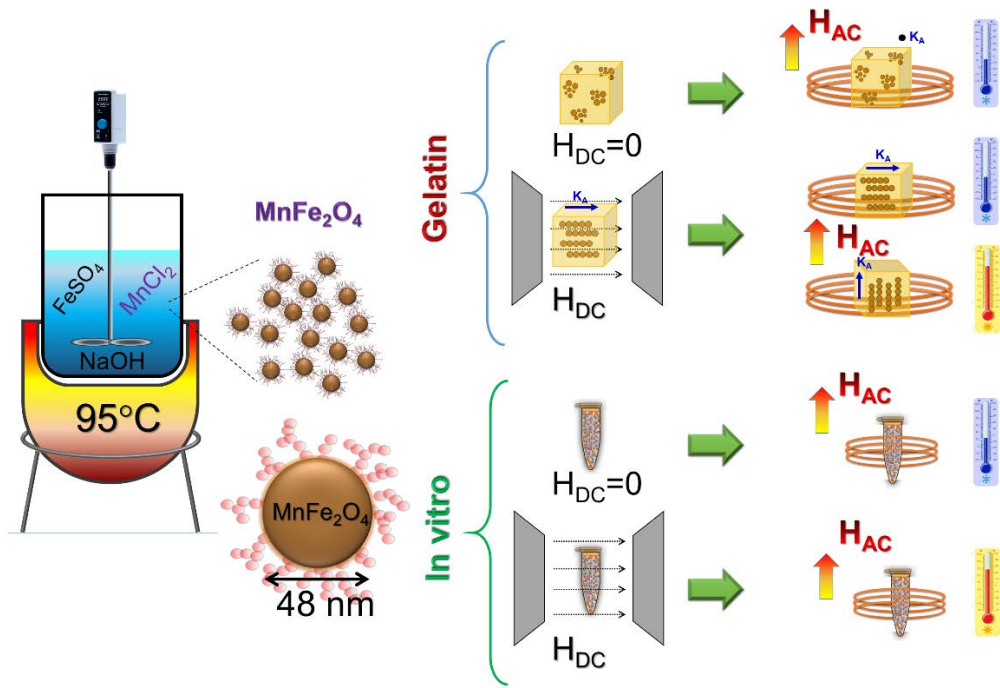

**Figure 1.** Schematic illustration of the synthesis procedure for the PEI-coated MnFe$_2$O$_4$ nanoparticles. The scheme also shows the procedure for inducing chain formation in phantoms and *in vitro* under an applied constant field H$_{DC}$ and the sample conditioning for power absorption measurements under different sample orientations.

### *2.1. Characterization of MNPs*

The characteristics of each sample, such as the morphology, median diameter (<d>) and standard deviation (σ), chemical composition and crystalline structure, were obtained with a 200 kV

TEM/STEM microscope working at 200 kV. HRTEM images were acquired using an XFEG TITAN (Thermo Fisher Scientific) operated at 300 kV and equipped with a monochromator and a probe aberration corrector (CEOS). The composition of the samples was extracted from EDX measurements on powder-pressed samples using a Nova Nanolab200 FIB-SEM microscope (Thermo Fisher Scientific) operated in SEM mode at 15 kV. Samples for HRTEM images were conditioned by placing a drop of the water-based colloids onto a holey carbon-coated copper microgrid and drying in air for 3-4 h. Electron diffraction patterns were indexed using a cubic spinel phase (sample card No. 22–1086 of the International Center for Diffraction Data® (ICDD®)). Particle size distributions were determined from TEM images at several grid positions, and the different size histograms acquired were fitted using a Gaussian distribution function $g(d)$ of the parameters ($d_0,\sigma$):

$$g(d) = A\, exp\left[-\frac{(d-d_0)^2}{2\sigma^2}\right] \qquad \text{Eq.1}$$

where A is the amplitude of the Gaussian, $d_0 = <d>$ is the mean value of the diameter distribution and $\sigma = \frac{FWHM}{2\sqrt{ln4}}$, with FWHM = full width at half-maximum. We used a Gaussian distribution to fit the data since it has been proven more adequate to represent the distribution of parameters of linear dimensions (i.e., particle diameter) as in the present case, whereas the log-normal functions are better suited for those parameters related to some power of a linear dimension (e.g. volume or surface area)[30]. From a practical perspective, we found that the goodness of fit was slightly better when the Gaussian fit was used instead of the log-normal function. The <d> and σ values were obtained from the fits to the particle size distributions in each sample were compared to provide an estimation of the synthesis reproducibility in terms of <d> and σ among different batches (see the Results and Discussion Section below). EDX analysis from TEM and SEM microscopy provided the atomic composition in terms of the ratio $R = \frac{[Fe]}{[Mn]}$ of metal ions, which is tabulated for different samples in Table S1 of the Supporting Information. The crystalline structure of the samples was evaluated using the intensity profiles extracted from selected area electron diffraction (SAED) analysis. The profiles showed patterns consistent with the spinel structure of $MnFe_2O_4$ (manganese ferrite). The indexing of the diffraction peaks, represented by the Miller indices at the top of the peaks, was consistent with the ASTM values and confirmed the crystalline structure. Additionally, the high-resolution images were subjected to Fourier transform analysis to identify the crystalline planes (data not shown). HRTEM images also showed the polymer coating present on the particles. For HRTEM analysis of MNP agglomerates, the MNPs were dispersed into an EPON™ (EPON 812) resin and sectioned with an ultramicrotome.

### 2.2. *Thermogravimetric Analysis (TGA)*

TGA of the powdered samples was performed using a TGA/DSC 1 (Mettler Toledo) from room temperature to 900 °C at a fixed heating rate of 10 °C/min under a continuous flux of nitrogen.

### 2.3. *Magnetic Measurements*

Magnetization M(T,H) measurements were performed on a commercial SQUID magnetometer (MPMSXL Quantum Design) on the *as prepared* colloidal samples, previously conditioned in plastic capsules as described elsewhere.[31] Magnetic measurements for quantification of MNP uptake in BV2 cells were performed at room temperature in a vibrating sample magnetometer (Lake Shore 7400 Series VSM) as a function of the field up to H = 1.5 T. The thermal dependence M(T) of the magnetization was measured using ZFC and FC protocols, in all cases with increasing temperature.

Ferromagnetic resonance (FMR) measurements were performed at 300 K in an Elexsys E500 spectrometer (Bruker) with X-band (9.4 GHz) bridge, applying the following parameters: Field

Sweep Width of 795 kA/m, attenuation of 20 dB, conversion time of 58.59 s and modulation amplitude of 795.77 A/m. For all measurements, the samples were conditioned by dispersing the MNPs on a polyacrylamide gel by sonication for 2 hours. This MNP-containing gel was dried along 8 hours under a dc magnetic field of 637 kA/m (in an air gap of 2 cm between two polar pieces of permanent magnets of diameter = 8 cm). After that, a much smaller area (dimensions of 0.5x0.2x0.2 cm$^3$) from the center of the solid polyacrylamide was extracted and placed at the FMR spectrometer sample space to measure the angular dependence of the resonance field $H_r$ with the applied field.

### 2.4. *Specific Power Loss (SLP) Experiments*

The power absorption experiments were performed in a commercial ac field applicator (DM100, nB nanoscale Biomagnetics, Spain), with the magnetic field frequency adjustable in seven steps within the range $229 \leq f \leq 831$ kHz, and the field amplitude $H_0$ variable within $7.95 \leq H_{AC} \leq 24.0$ kA/m. Measurements were made using an RF-immune temperature sensor to obtain the time dependence of the sample temperature. The SLP values of as-prepared samples were extracted from the initial 20-40 seconds of heating curves using a linear fit of this region (differences with $\Delta T/\Delta t$ values from nonlinear fits were in all cases less than 2%). In addition to experiments on the *as-prepared* colloids, SLP measurements were performed on MNPs embedded in a gelatin matrix that blocks Brownian rotation of the particles to mimic the cytoplasmic environment of *in vitro* experiments. For this step, known amounts of MNPs from colloids were dispersed into microbiological-grade (>97% collagen) gelatin with high wettability and solubility, forming a water-based compound composed of hydrolyzed collagen that can be applied at high concentrations in a liquid state at a relatively low temperature (T = 90 ºC) before solidifying at room temperature into a solid matrix. The SLP values of both *as-prepared* colloids and gelatin-embedded samples containing a total mass $m_{NP}$ of MNPs were calculated using the initial slope of the T *vs*. t curve (known as the maximum derivative method), with the following expression:

$$SLP = \frac{m_l c_l + m_{NP} c_{NP}}{m_{NP}} \left(\frac{dT}{dt}\right)_{max} \qquad \text{Eq.2}$$

where $c_l$ and $c_{NP}$ are the specific heat capacities of the liquid carrier (or gelatin matrix) and the MNPs, respectively. The temperature increase *dT* in the sample measured in the time interval *dt* was extracted numerically from the *T(t)* data, and the maximum derivative of the initial temperature increase (within the first 50-100 seconds of the experiment) was taken as the 'adiabatic' approximation of the heating process. In all experiments, the concentration of MNPs was taken to be ≲ 2% wt. and therefore the term $m_l c_l + m_{NP} c_{NP} \approx m_l c_l$ in the previous equation was approximated by the following:

$$SLP = \frac{c_l \delta_l}{\phi}\left(\frac{\Delta T}{\Delta t}\right) \qquad \text{Eq. 3}$$

where $\delta_l$ and $\phi$ are the density of the liquid and the (mass) concentration of MNPs in the colloid, respectively.

### 2.5. Cell culture and MNP uptake experiments.

Cell culture of BV2 cells (ATCC CRL-2469) was performed using Dulbecco's modified Eagle's medium (DMEM) with Ham's F12 medium (1:1) and 15% fetal bovine serum supplemented with 2 mM L-glutamine, 100 IU/mL penicillin and 100 mg/mL streptomycin. The growth of the cells was controlled in constant-temperature incubators at 37 °C in a mixture of air and $CO_2$ at a 95/5% composition. The SLP experiments in cell pellets were performed after the cells were incubated for 24 h with different concentrations of MNPs and then washed several times, and the culture medium was replaced by fresh, ordinary DMEM medium. Cells for control samples (i.e., without MNPs) were grown simultaneously in each experiment to assess the fluctuations in both intrinsic and extrinsic parameters of the cell pellets among the different experimental runs.

### *2.6. Cell Viability Assays.*

To assess the cell viability after incubation with MNPs, ≈3-8x$10^5$ cells were seeded into a 12-well plate and incubated for 24 h at 37 ºC with 5% $CO_2$. The culture medium was then replaced with increasing MNP concentrations, and the plates were incubated for 24 h. Trypan blue assays were then conducted by diluting 20 mL cell samples into Trypan blue (1:1), and the viable cells were counted. The % cell viability was calculated based on the control and 100% viability.

### *2.7. Intracellular Distribution of MNPs*

For the study of the intracellular MNP distribution by TEM and FIB-SEM, the samples were prepared by seeding cells (1×$10^6$ cells/well) in a 6-well plate in 2 mL of culture media. The next day, different concentrations of MNPs were added and incubated overnight. After incubation, the cells were washed, detached and fixed with 2% glutaraldehyde solution for 2 h at 4 °C. The cells were then washed three times in cacodylate buffer (pH 7.2) and treated with 2.5% potassium ferrocyanide and 1% osmium tetroxide for 1 h at room temperature. After washing, the cells were dehydrated with increasing concentrations of acetone (30% (x2), 50% (x2), 70% (x2), and 90% (x2)), followed by further dehydration with 100% acetone. After drying, the samples were embedded in a solution (50:50) of EPON resin and acetone (100%) overnight and then for 4-5 h in 100% EPON resin. These samples were maintained for 2 days at 60 °C.

### *2.8. Dual-Beam FIB-SEM Analysis of Intracellular Agglomerates*

Three-dimensional reconstruction of the intracellular clusters of MNPs was achieved by sequential cross-sectioning of the cells with a dual-beam FIB/SEM (Nova NanoLab 200 and Helios 650, Thermo Fisher Scientific). Prior to these observations, the BV2 cells were fixed, contrasted and embedded in resin after co-cultivation with MNPs, following the protocol described above. SEM images were taken at 2–5 kV with a field emission gun (FEG) column. The cross sections of single cells were performed with a Ga beam at 30 kV with currents between 10 pA and 20 nA. Energy dispersive X-ray (EDX) spectra were acquired to confirm those spots with Mn and Fe signals to assess the presence of MNPs.

## 3. Results and Discussion

### 3.1. *Characterization of the MNPs and aggregates*

The analysis of the morphology, composition and structure of $MnFe_2O_4$ MNPs was performed using different electron microscopy techniques. Transmission electron microscopy (TEM) images showed that all samples were composed of partially faceted MNPs with definite shapes, as shown in **Figure 2**a (see also Figure S1 in the Supporting Information). A statistical analysis of many colloidal samples from n = 5 different batches was performed regarding the average particle size <d> and size distribution σ of the resulting MNPs. Fitting the histograms with a Gaussian distribution yielded average sizes of $45 \leq \langle d \rangle \leq 51$ nm and standard deviations of $11 \leq \sigma \leq 15$ nm for the different batches characterized (Figure S2 in the Supporting Information). We constructed a parameter designed to provide a rough quantitative estimate concerning the reproducibility of this synthesis method by comparing the average $\langle d \rangle$ values among different batches and taking the maximum difference $\Delta \langle d \rangle = max|\langle d \rangle_i - \langle d \rangle_j|$ between all pairs of values. A maximum deviation of $\Delta \langle d \rangle \approx 6$ nm was found among all samples, implying that all distributions have $\langle d \rangle$ values within a one-sigma deviation and therefore all batches can be treated as indistinguishable. The physical parameters are summarized in Table 1.

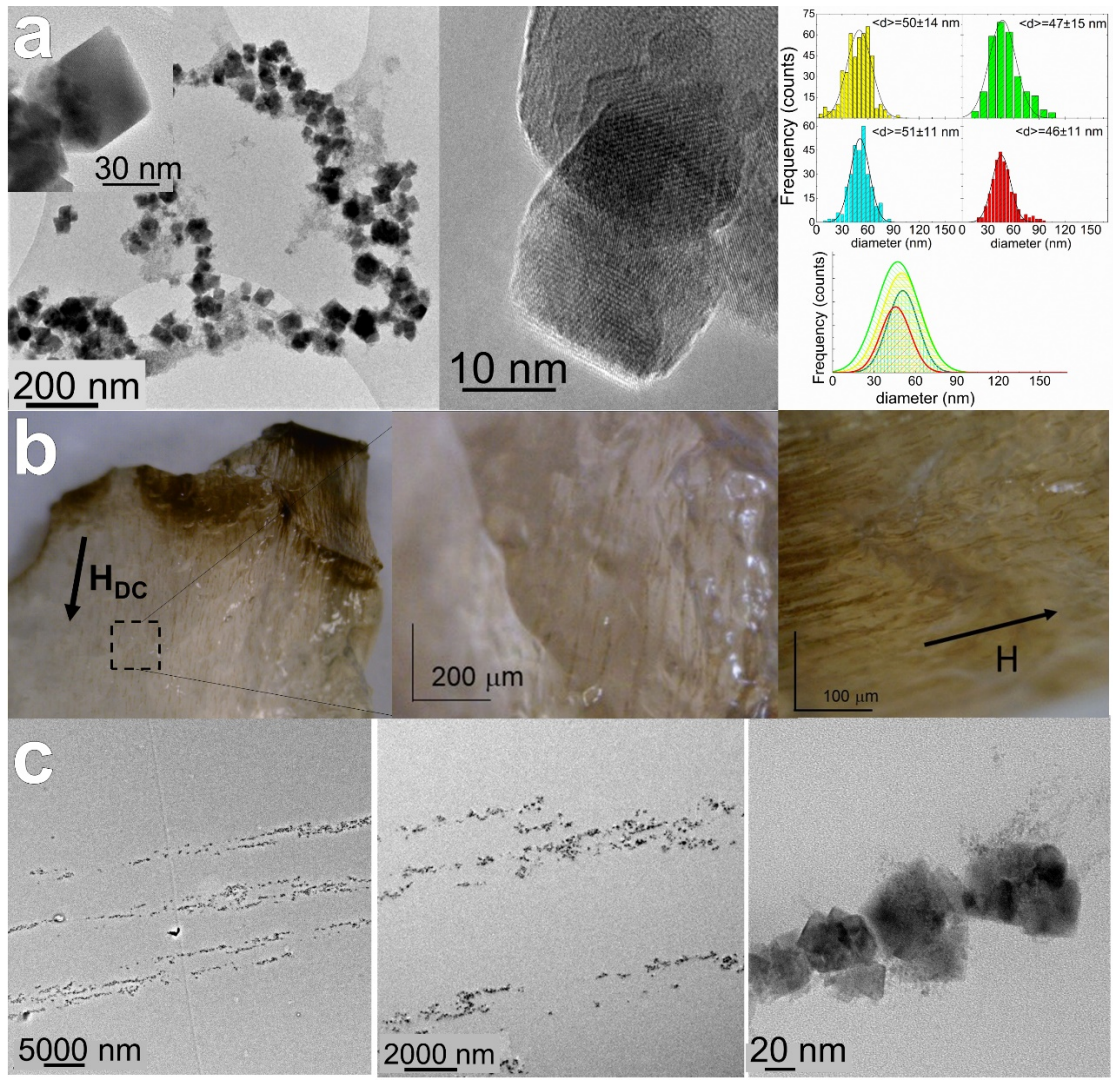

**Figure 2:** a) Transmission electron micrographs of *as-prepared* MnFe$_2$O$_4$ MNPs and the resulting size distribution from four different batches. b) Optical microscopy images of chain-like clusters obtained in gel phantoms solidified under a dc magnetic field: H$_{DC}$ = 1.42×10$^3$ kA/m (17.8 kOe) and c) TEM images of the same chains obtained under H$_{DC}$ showing the continuance of the chain structure at the nanometer scale.

Analysis of energy dispersive spectroscopy-scanning transmission electron microscopy (EDX-STEM) and selected area electron diffraction-high-resolution TEM (SAED-HRTEM) data at the single-particle level was performed from C$_S$-corrected HRTEM images. The data could be indexed with the spinel structure, Fd3m space group, and unit cell parameter a = 8.5111 Å. The data showed no evidence of distortions, crystal defects, or any preferential orientation of the nanoparticles. Fast Fourier transform (FFT) analysis in the $[1\bar{1}\bar{2}]$ zone axis showed spots corresponding to $(\bar{4}0\bar{2})$, $(2\bar{2}0)$, (042) and (222) planes of the cubic spinel crystal structure. A thin layer of 1-3 nm on the MNP surface can be observed in the HRTEM images, corresponding to the polyethylenimine (PEI) polymer layer added *in situ* during synthesis, which has the double function of providing a surface coating on the particles and controlling the final size, as previously reported elsewhere.[32,33,34]

**Table 1.** Typical Composition and Magnetic Parameters of the MnFe$_2$O$_4$ MNPs from the different batches studied in this work (n = 4). The nominal atomic compositions (Mn$_x$Fe$_{3-x}$O$_4$) were obtained from EDS-SEM and EDS-STEM. The average particle diameter ⟨d⟩ and standard deviation σ were obtained from Gaussian fits of each of the particle distributions. The saturation magnetization $M_S$ and coercive field $H_C$ at T = 10 K and 300 K are also tabulated.

| Sample | ⟨d⟩ (nm) | σ (nm) | T = 10 K | | T = 300 K | |
|---|---|---|---|---|---|---|
| | | | $M_S$ (Am²/kg) | $H_C$ (kA/m) | $M_S$ (Am²/kg) | $H_C$ (kA/m) |
| Mn$_{0.97}$Fe$_{2.03}$O$_4$ | 46 | 15 | 67.94 | 17.54 | 43.21 | 0 |

The global stoichiometry of the MNPs regarding manganese and iron contents was assessed by collecting EDX spectra in different macroscopic regions of the samples. This analysis was performed by averaging scanning electron microscopy-EDX (SEM-EDX) spectra over a minimum of five areas from each sample (Figure S3 in the Supporting Information). Since the electron beam has a spot of ≈10-50 nm and the collected X-rays have a penetration depth of approximately 1–3 µm, the atomic composition corresponds to a sample volume of ≈0.2 µm³. The chemical composition was further analyzed through STEM-EDX spectroscopy for few-particle clusters and at the single-particle level (see Table S1 in the Supporting Information). Similar Fe:Mn ratios were derived for both individual MNPs and few-particle clusters, indicating a homogeneous chemical composition of the MNPs. Typical magnetic parameters of the MNPs in this work are summarized in Table 1, showing a nearly stoichiometric atomic manganese(II) and iron(II) composition, with small variations among different batches; the largest deviation observed was $Mn_{0.97}Fe_{2.03}O_4$, i.e., a slightly Mn-defective manganese ferrite. The presence of the PEI coating was confirmed by thermogravimetric analysis (TGA), which showed the existence of a first mass loss at approximately 420 K related to the decomposition/carbonization and partial detachment of the PEI coating (Figure S4 in the Supporting Information). A second temperature range of mass loss centered at ≈ 840 K is likely related to evaporation of the residual carbonized organic material starting at T ≈ 800 K. The analysis also provided the mass percentage of the organic PEI coating (≈3%), consistent with the organic layer of t ≈ 2 nm thickness observed in HRTEM images. This result was used to correct the saturation magnetization $M_S$ by the mass of the inorganic phase.

The magnetic data of the *as-prepared* colloids showed that the MNPs were blocked at room temperature, as inferred from the irreversibility between the zero-field cooling (ZFC) and field cooling (FC) magnetization curves in the whole temperature range up to room temperature (Figure S5 in the Supporting Information). Hysteresis loops recorded at T = 10 K provided a magnetic saturation value $M_S$ = 67.94 Am²/kg, which decreased to $M_S$ = 43.21 Am²/kg at room temperature. The thermally assisted change in $M_S$ is linked to changes in the coercive field $H_C$ by the relation $H_C(T) = \frac{2K_u}{\mu_0 M_S}$, being $K_u$ the magnetic anisotropy constant and $\mu_0$ the magnetic permeability of vacuum. Using this relation we obtained a $K_u$ (10 K) = 8.1 kJ/m³ and $K_u$(300 K) = 0.21 kJ/m³, a change by a factor ≈40 between these temperatures. Taking into account the value at room temperature a blocking diameter of D = 97 nm is obtained. Additionally, using the thermal dependence of $M_S$ and $K_u$ the obtained critical diameter for the single-domain transition was D=80 nm. Although these numbers were obtained from very simplified models, the general conclusion is that the size dispersion in the present samples implies the coexistence of MNPs in both the SPM and blocked states.

Our attempts to fit the thermal evolution of the coercive field $H_C(T)$ with the expected thermally activated Néel-Arrhenius magnetic relaxation expression for single-domain MNPs [31] failed due to evident differences in the low- and high-temperature $H_C(T)$ dependence. Including the Brukhatov-Kirensky relation [35] to account for the thermal dependence of $K_{eff}$ was also unsuccessful (see the Supporting Information and Figure S6 for details on the fitting procedure). These deviations suggest that the MNPs are not single-domain particles but possess a more complex magnetic structure, since above and close to the single- to multi-domain transition, a non-linear spin configuration should provide a minimum total energy while keeping the saturation magnetization close to the bulk value [36,37]; it has consequently been proposed to use MNPs with sizes close to the single- to multidomain transition to improve local heating in biological media.[38,39]

Based on the above considerations, we performed ferromagnetic resonance (FMR) experiments to measure the particles' magnetic field anisotropy and estimate an effective magnetic anisotropy constant $K_{eff}$ of the oriented nanoparticles. To obtain this value, the angular dependence of the resonance field $H_r$ of the FMR lines in these particles was used to extract $K_{eff}$ through the following relation:

$$H_r = \frac{\omega}{\gamma} - \frac{3}{2} H_A^{eff} [cos^2(\theta) - 1/3] \qquad \text{Eq. 4}$$

The resonance field approximately follow the expected $H_r \propto cos^2\theta$ angular dependence (Figure S7 in the Supporting Information), although the details of the $H_r(\theta)$ function indicates minor contributions from higher-order anisotropy terms, suggesting a more complex magnetic structure than a uniaxial single-domain core. Before FMR measurements, the samples were oriented following the procedure described in the Materials and Methods section. Thus, it should be noted that the procedure can produce chain structures, affecting the effective anisotropy values of isolated MNPs, as recently demonstrated.[40] Assuming a 'perfect' single-domain magnetic structure of the MNPs, the fit of the experimental data with Eq. 4 yielded an anisotropy field value $H_A^{eff} = 319 - 115\ Oe$. Using a density value of δ= 5020 kg/m$^3$ for the MnFe$_2$O$_4$ spinel[41] and the experimental saturation magnetization values μ$_0$M$_S$ = 43.2-56.1 Tm$^3$/kg at T = 300 K (from the M(H) data), we estimated an effective magnetic anisotropy value within K$_{eff}$ = 2.6 - 3.5 kJ/m$^3$ at T=300 K This value is somewhat lower than the corresponding magnetocrystalline anisotropy of bulk MnFe$_2$O$_4$, supporting our hypothesis about deviations from a single-domain magnetic structure in the MNPs mentioned above. It should be mentioned that the relaxation measured by FMR experiments is based on the system being excited at a local energy minimum and thus the curvature of the actual energy barrier is not known. The use of an 'effective' anisotropy constant K$_{eff}$ can be a limited approximation of the real energy barrier in magnetic relaxation modeling, depending on the experimental techniques used.

### 3.2. *Specific Loss Power of MNP Aggregates in Gelatin Phantoms and Cells*

As shown in **Figure 3**, the SLP of the *as-prepared* MnFe$_2$O$_4$ colloids was 727±14 W/g for an applied field H$_0$ = 24 kA/m and a frequency *f* = 571 kHz, the standard values chosen for frequency and field of all SLP data in this work, unless otherwise stated. In the colloidal state, contributions from both Néel and Brown relaxation mechanisms are expected since the MNPs are able to rotate more or less freely in the liquid carrier. The SLP values of the *as-prepared* colloids as a function of $H_0$ showed a power-law dependence (SLP($H_0$) = A $H_0^\lambda$) with $H_0$, and the best fit using this expression yielded λ=1.9 (see Figure S8 in the Supporting Information). Interestingly, this value is close to the λ = 2 value predicted by the linear response theory (LRT) [6, 29] although this model is not valid regarding the condition $H_0 \ll H_K$ (i.e., the applied field $H_0$ be much lower than the anisotropy field $H_K$ of the MNPs), which is clearly not satisfied by the present low-anisotropy MNPs since $H_0$= 24 kA/m and $H_K$ = 25.4 kA/m. We do not yet have a clear explanation for this field dependence.

As mentioned in the Introduction, the current consensus is that the SLP values measured during *in vitro* or *in vivo* experiments are lower than the values for corresponding *as-prepared* colloids made from the same nanoparticles. How different these SLP values are between the two situations depends primarily on the magnetic dipolar interactions and the magnetic anisotropy of the MNPs but also on the physicochemical conditions in the biological media or the cell cytoplasm, mainly the actual viscosity at the location of the MNPs. To mimic the cytoplasmic environment of *in vitro* experiments, we performed systematic SLP measurements in gel phantoms by dispersing the MNPs into microbiological-grade (>97% collagen) gelatin, a water-based compound composed of hydrolyzed collagen that can be used at high concentrations in liquid state at relatively low temperatures (T = 90 °C) before solidification into a solid matrix at room temperature. The MNPs dispersed in gelatin were conditioned inside cylindrical sample holders and allowed to cool to room temperature in an applied field of zero. For the formation of MNP chains, the same procedure was followed except that immediately after adding the MNPs, they were cooled to room temperature under an applied field ($H_{DC}$ = 1.42x10$^3$ kA/m (17.8 kOe)).

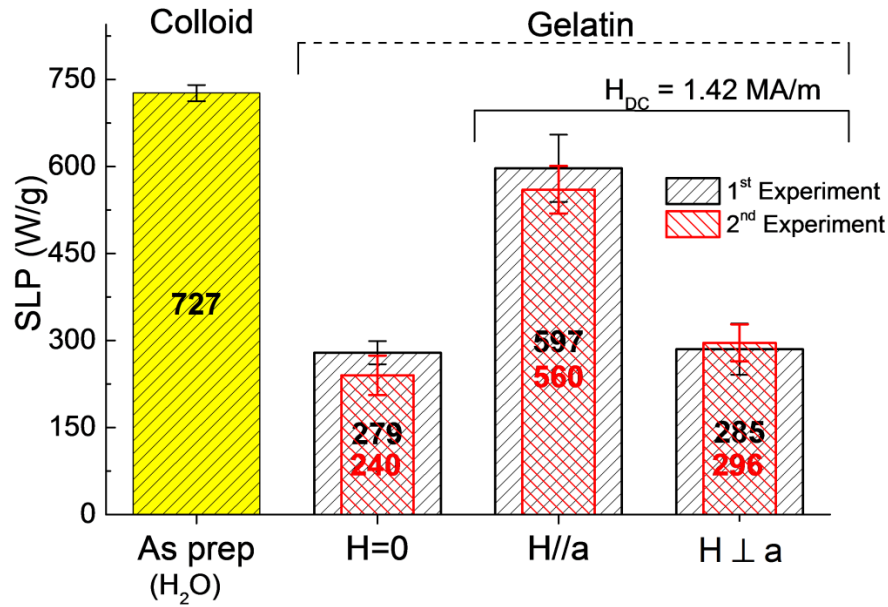

**Figure 3.** SLP values ($H_{AC}$ = 24 kA/m; $f$ = 571 kHz) for as-prepared MnFe$_2$O$_4$ colloids measured in water (as prep sample) and in gel-based phantoms. The H=0 bars indicate the SLP values obtained from the gel-based phantom containing MNPs (center) with solidification without any applied field. The H//a and H⊥a bars are the SLP values for tthe magnetic chains formed under $H_{DC}$ = 1.42x10$^3$ kA/m when the chains are oriented, measured with their major axis parallel (H//a) and perpendicular (H⊥a) to the applied ac magnetic field $H_{AC}$.

Optical microscopy analysis of the samples solidified under $H_{DC}$ = 0 showed that only agglomerates of sizes ≲ 200 nm were present. In contrast, for the MNPs embedded in gelatin and solidified under a magnetic field $H_{DC}$ showed the formation of elongated clusters with lengths of a few tens of microns (see **Figure 2**b). Despite the homogeneity of the field within the sample space (<3% within 10 × 8 × 8 mm$^3$) applied during chain formation, the images revealed some migration of the MNPs to both magnetic poles, resulting in a nonhomogeneous MNP distribution within the sample. Taking advantage of this situation, we analyzed several sample regions and observed that in areas of lower concentration, the length and width of the clusters were smaller, suggesting that these dimensions are determined by the local concentration of MNPs. This conclusion is consistent with previous reports of concentration-dependent chain dimensions observed in several ferrofluids[42,43] and polymers[44]

Experimental and theoretical evidence on MNP chain formation under applied ac or dc magnetic fields dates back to the 1970s [42-43, 45]. These and subsequent works have confirmed that the length of the formed chains, from a few nanometers to the micrometer range, increases with larger applied fields. These pioneering works were based on indirect measurements of optical polarized light, and therefore, the researchers were unable to directly observe smaller units; one is tempted to speculate that nanometer-length chains were also present but undetected. A recent report on self-assembled $Fe_3O_4$ chains in the gas phase has demonstrated that magnetic dipole-dipole interactions are the driving force behind chain formation and that the details of their magnetic anisotropy easy axis along the [111] crystallographic direction also play a role [46]. However, in highly viscous liquid phases, viscous drag dominates the kinetics of cluster formation by rescaling the aggregation time, although dipole-dipole interactions between particles remain the physical mechanism behind this process [47]. Our analysis using TEM images confirmed the nearly 1D character of the chains formed under $H_{DC}$ magnetic fields, as shown in **Figure 2**, with lengths from ≈100 nm to several microns (see also Figure S9 in the Supporting Information).

Power absorption measurements of the aligned chains were performed with samples oriented with their major axis parallel and perpendicular to the applied ac magnetic field (**Figure 3**). The obtained data unambiguously showed an increase by a factor of ≈2 in SLP when the chains were oriented parallel to the applied ac magnetic field with respect to the sample solidified under $H_{DC}$

= 0. **Figure 3** also shows the angular dependence of the chain axis orientation on the direction of the applied ac magnetic field: rotating the chain's axis by 90 degrees yielded a decrease in SLP by a factor of ≈2. These findings are similar to a previously reported [48] SLP study on previously oriented MNPs embedded in a solid matrix, particularly regarding the decrease in SLP in the perpendicular configuration to values similar to those in the $H_{DC} = 0$ cluster configuration. The increased SLP observed after chain formation, which was greater than the experimental error, is related to the Néel relaxation mechanisms in these systems since the contribution to the SLP from Brownian relaxation was eliminated in the gelatin matrix. As will be discussed below, numerical calculations using a simple mean-field model reproduced the SLP values for clusters and linear chains.

### 3.3. In Vitro *Analysis of Magnetic Agglomerates*

Before running the SLP experiments *in vitro*, the toxicity of the $MnFe_2O_4$ MNPs was assessed on BV2 cell cultures with increasing concentrations up to 100 μg/mL, resulting in cell viability values ≥95% at the maximum MNP concentration. Furthermore, we assessed the impact of cell stress under the experimental conditions of in-field cell culture by placing a control cell culture under identical conditions (2-3x10$^6$ cells in 1 mL of DMEM in an Eppendorf tube for 15 h), obtaining viabilities of 89-98% after incubation (see Figure S10 in the Supporting Information).

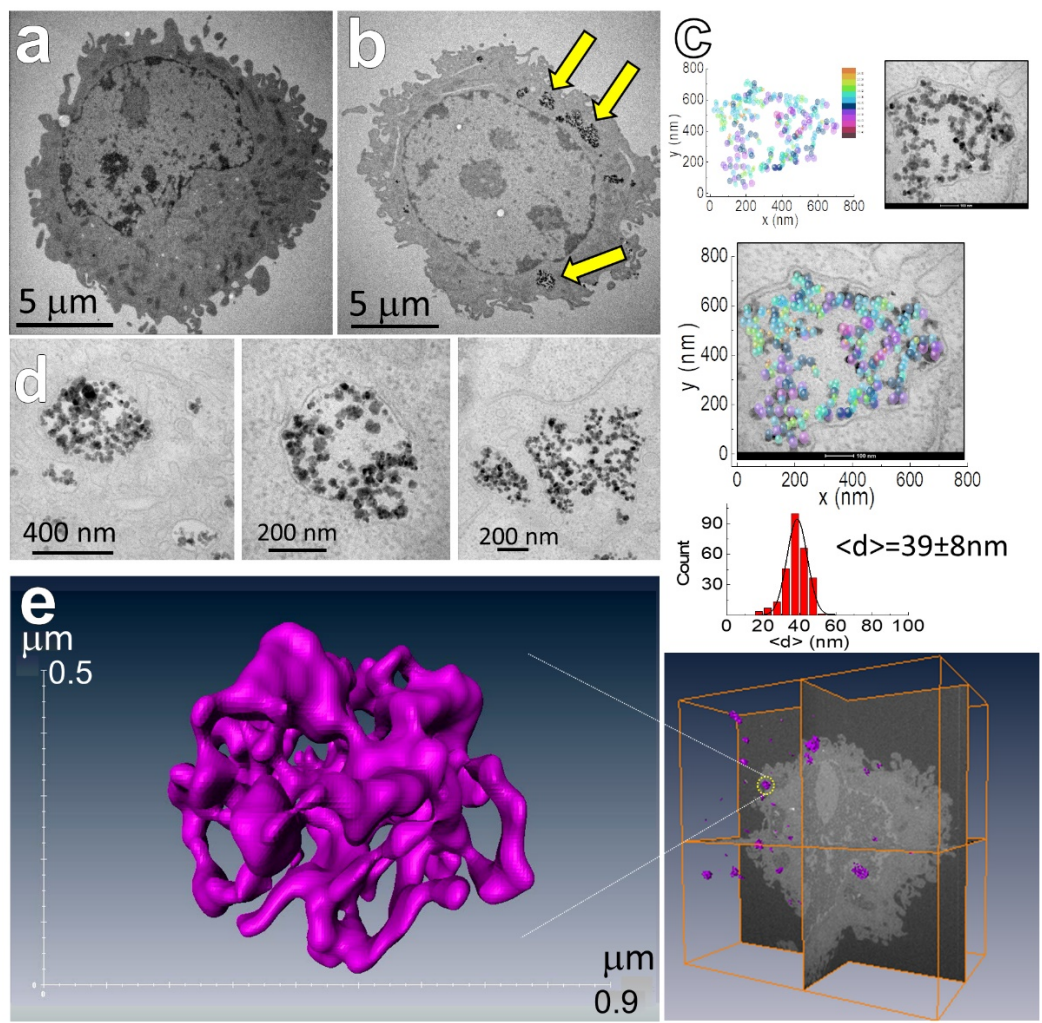

**Figure 4.** *In vitro* cluster analysis from TEM and FIB images. a) TEM images of control BV2 cells without MNPs and b) cells cultured overnight with 50 μg/mL $MnFe_2O_4$ showing MNP clusters (arrows). c) TEM 2D image of a selected agglomerate, the reconstruction of the MNPs coordinates and the superposition of both experimental and reconstructed MNP distributions. The intracellular particle size distribution is also shown. d) Magnification of three different BV2 cytoplasmic regions showing variable cluster structures and shapes. e) Snapshot of the three-dimensional distribution of MNPs as reconstructed from FIB slicing of a single cluster (a render animation is available in the file MovieH=0.mp4 of the Supporting Information).

In order to obtain detailed data on the geometry of the clusters formed 'naturally' (i.e., $H_{DC} = 0$ during incubation) within the cell cytoplasm, we performed a series of 'standard' *in vitro* uptake experiments using BV2 microglial cells. The protocol was standardized by fixing the amount of MNPs (50 μg/mL) and the time of overnight cell culture (15-18 h). After incubation, the cells were washed several times to eliminate the excess unattached nanoparticles and then dehydrated and fixed using EPON resin as described in the Experimental Section. Both TEM and dual-beam focused ion beam-SEM (FIB-SEM) techniques were used to analyze the MNP clusters and for comparison with control cells (i.e., without MNPs). The images showed no perceptible changes in the morphology of the BV2 cells after particle uptake (**Figure 4** a and b). Typically, the distribution of MNPs within the intracellular space of BV2 cells resulted in large agglomerates (arrows in **Figure 4**b) composed of many hundreds of MNPs. The particle size distributions obtained from several (n = 7) intracellular agglomerates after uptake yielded distribution values $\langle d \rangle$ from 39±8 nm to 47±9 nm, similar to the *as-prepared* colloid distributions (**Figure 4**c). This similarity between the size distributions of as prepared colloids and intracellular clusters suggests that, irrespective of the cell pathway involved, for the type of MNPs and BV2 cells studied in this work this uptake mechanism is no size-selective.

Considering the 2D projection nature of the observed TEM images, it can be inferred that the actual 3D agglomerates should be made of many hundreds to a few thousand particles, in agreement with previously reported data [49]. The actual three-dimensional geometry of the clusters was obtained from 3D reconstruction of selected BV2 cells using dual-beam SEM-FIB microscopy by slicing a single cell (≈400-500 slices) and reconstructing the cluster structure through AMIRA® software. The reconstruction shown in **Figure 4**e (see also the 3D animation file ClusterH=0.mp4 in the Supporting Information) clearly confirmed a global spherical symmetry of these clusters inside the cell cytoplasm, with a total volume of MNPs (i.e., the colored space in **Figure 4**e) ($V_{MNP}$) of 0.265 μm$^3$, suggesting that this cluster comprises an average number of ≈7700-8300 MNPs. Considering that this selected cluster was among the smallest clusters observed within BV2 cells and assuming a similar average particle density within all clusters, a more realistic number for clusters with an average volume (i.e., ≈2-4 μm$^3$) should be closer to ≈10$^5$ MNPs. We note here that the actual size and shape of intracellular clusters depend on the characteristics of the MNPs (e.g., surface and size) and the cell line used in any specific experiment. The PEI-coated MNPs used in this work have previously been reported to localize in the cytoplasmic space of BV2 cells, without any noticeable amount of particles on the cell membrane [28]. The same characteristics were observed in the present work.

The magnetic response of the aggregates within the BV2 cells was essentially identical to those from the *as-prepared* colloids. Both the M(H) and ZFC-FC curves (data not shown) and the $H_C(T)$ data taken from 5 K to room temperature showed slightly larger anisotropy values than those for the *as-prepared* particles (Figure S6 in the Supporting Information). Attempts to fit the data using a model of single-domain MNPs also failed for these aggregates, suggesting a non-uniform spin structure of the nanoparticles and the presence of dipolar interactions, as discussed above.

### 3.4. In Vitro *Study of Elongated Aggregates*

The biologically active mechanisms of cells for transporting MNPs or storing MNPs in endosomal or lysosomal structures promote agglomeration within these vesicles, which in turn favors demagnetizing arrangements within the agglomerates, decreasing the heating efficiency of the MNPs, which can be observed both theoretically and experimentally [39, 49].

An ingenious approach recently reported by Jeon et al. [49] provided an accurate description of the structure of MNP agglomerates in cell media, and their methodical investigation using image analysis showed that these objects have fractal dimensions $D_f < 3$. The authors stated that a lower fractal dimension implies that the *local* coordination of MNPs within aggregates is lower than expected for three-dimensional agglomerates, a key observation for theoretical SLP modeling. Unfortunately, the authors did not perform direct *in vitro* SLP measurements that would allow corroboration of the accuracy of the approach. Similar considerations apply to the recent work by Liu et al. [50] on *in vivo* application of magnetic nanorings that provided an insightful approach that demonstrated an increase in $CD8^+$ induced by mild hyperthermia and the inhibition of the immunosuppressive response of the tumor, as well as an extensive physical characterization of the nanorings, including SLP data on gel phantoms. Unfortunately, no direct *in vitro* SLP data were reported to assess how these structure respond to the ac magnetic fields at the intracellular level.

Indeed, most of the results reported about *in vitro* or *in vivo* heating efficiency of MNPs are usually drawn from physiological parameters, such as cell death, reactive oxygen species, and metabolic activity, and virtually no direct SLP data are reported. Although the analysis of MFH experiments on a physiological basis is clearly satisfactory to assess clinical efficacy, such analysis cannot provide information about the actual heating efficiency of the MNPs employed (i.e., the SLP values under given conditions), which is needed to define *a priori* the concentrations needed for a given tumor volume and type. The physiological outcome of *in vivo*/*in vitro* hyperthermia experiments (except intrinsic cytotoxic effects) depends only on the exposure temperature and time, and similar outcomes could be, in principle, produced by any nanosystem capable of delivering similar amounts of power. To date, there are very scarce SLP data for *in vivo/in vitro* experiments that include a comprehensive physicochemical and magnetic characterization of nanoheaters for comparison with numerical models.

Image analysis of the particle agglomerates after in-field incubation of the MNPs (**Figure 5**) showed alignment of the clusters along the field direction, irrespective of cell borders or cell organelles. From the particle size histograms extracted from several of these clusters, we found distributions coincident with the *as-prepared* particle size distribution within the experimental error (see **Figure 5**b and c), indicating that there are no size-selective uptake mechanisms. Compared with the chains formed in resin, the agglomerates formed within BV2 cells have a less pronounced 1D shape. However, the three-dimensional reconstruction from dual-beam SEM-FIB microscopy shown in **Figure 5** (see also the 3D FIB rendering in Movie1.mpg and Movie2.mpg of the Supporting Information) clearly displays a clear alignment of the long axis of symmetry in the direction of the applied field.

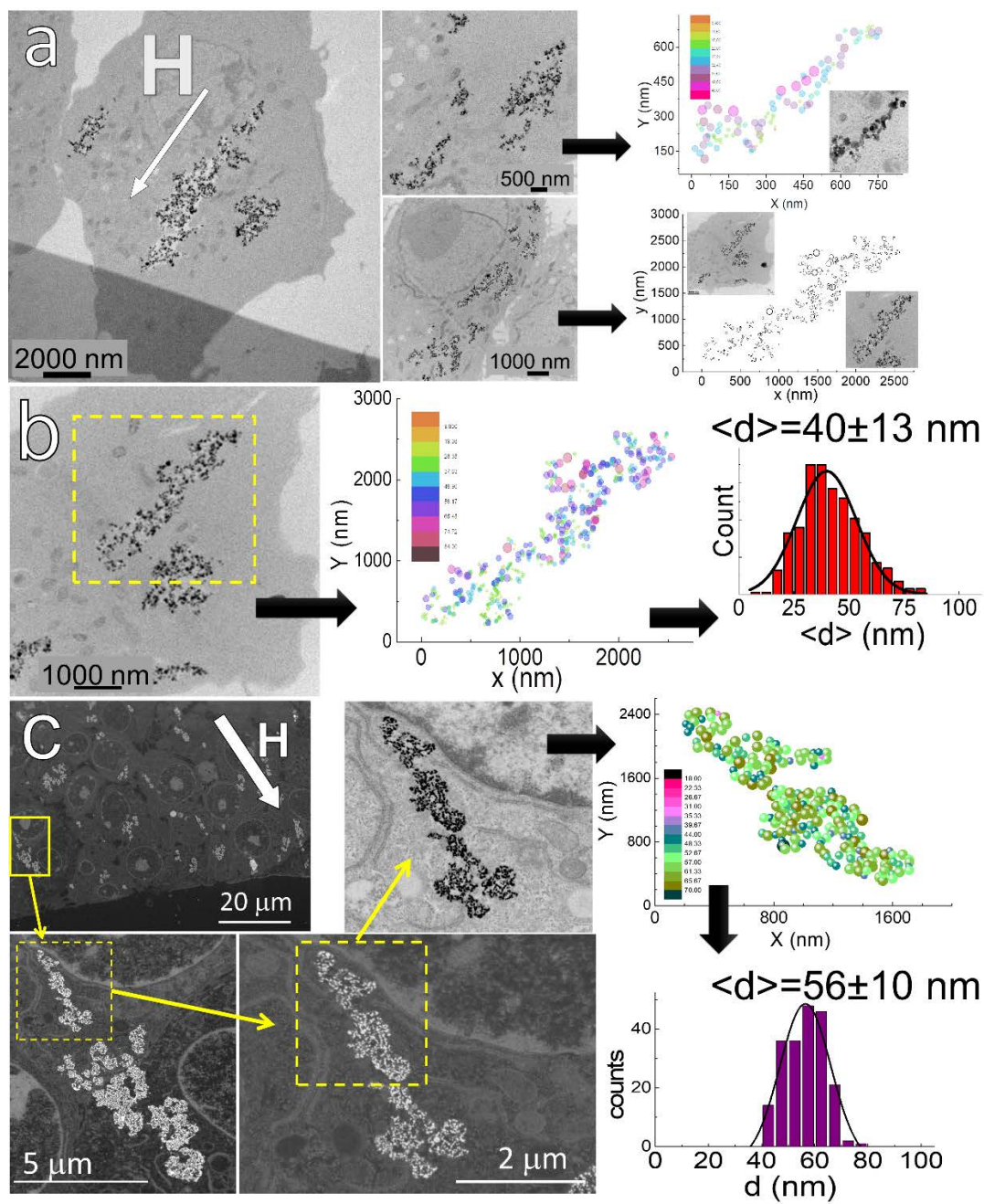

**Figure 5.** a) Bright-field TEM images of MNP clusters within BV2 cells grown under an applied field of $H_{DC}$ = 650 kA/m for 18 h. The selected areas for reconstruction of MNPs spatial and size distribution within the clusters is shown in the right panels. b) Same as (a) but including analysis of the resulting size distribution histogram and average MNP size of the cluster. c) Pathway for the selection and reconstruction using the SEM-FIB cross section of cells. The selected

agglomerates within BV2 cells grown under a magnetic field were selected from an appropriate snapshot image, the spatial distributions digitalized and the resulting MNP size distribution obtained. The average sizes obtained for the intracellular MNPs were coincident, within error, with the size of the MNPs in the *as prepared* colloids (for the corresponding 3D rendering of these clusters see Movie1.mpg and Movie2.mpg animation files in the Supporting Information).

### 3.5. *SLP of Spherical and Elongated* in vitro *Aggregates*

The reproducibility of *in vitro* SLP experiments is affected by the intrinsic cell variability in MNP uptake. For the present SLP experiments, a possible additional effect on the cells could arise from minor inhomogeneities in the magnetic field $H_{DC}$ applied during incubation, because magnetic field gradients can change the configuration of preexisting intracellular magnetic structures. We attempted to minimize this variability between experiments by repeating the SLP measurements in three independent cell cultures grown several weeks apart, with a very similar number of total cells each time, while keeping all experimental conditions (incubation time, MNP concentration, field application time, etc.) as constant as possible. In each of the three experimental runs, we performed SLP measurements in triplicate.

In spite of some interesting theoretical and experimental works on the effect of chain orientation to enhance the heating efficiency,[51] no attempt to reproduce these effect in vitro have been reported so far. The SLP experiments *in vitro* reported here were performed under the same experimental conditions than those used for the gel-based phantoms ($H_0$ = 24 kA/m; $f$ = 571 kHz) on two different samples: one sample was co-cultured overnight with an MNP concentration of 50 μg/mL, without any magnetic field applied, and a second sample was cocultured overnight under a constant dc magnetic field of $H_{DC}$ = 650 kA/m, with the same 50 μg/mL MNP concentration. In the case of cells incubated with MNPs under a constant magnetic field $H_{DC}$, we used an applicator designed to maintain the homogeneity of the field within 2% across the whole sample space. It is important to mention that during the SLP measurements, the cells are free to rotate in the culture medium within the sample holder, resulting in some degree of randomization of the MNPs chain axes relative to the $H_{AC}$ field direction.

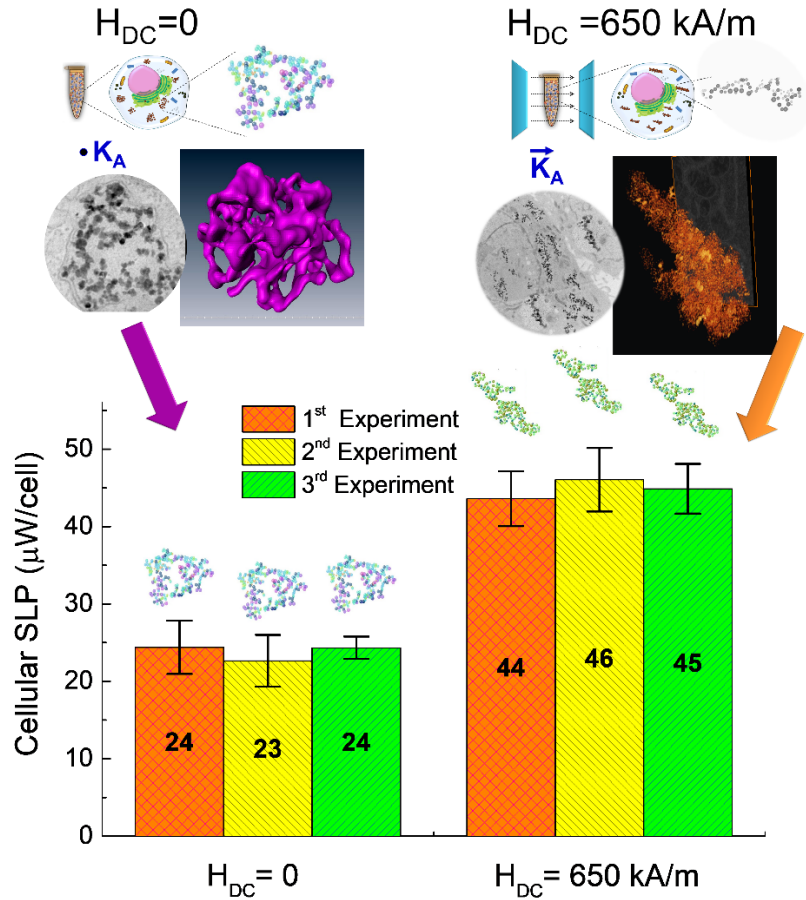

**Figure 6.** *In vitro* power absorption experiments for MNP clusters within BV2 cells formed overnight without an applied field ($H_{DC} = 0$) and under an applied dc field $H_{DC} = 650$ kA/m. SLP values are given per cell in µW/cell (n = 3).

The resulting SLP values expressed in µW per cell (**Figure 6**) showed a clear increase in SLP in the low-dimensional agglomerates relative to that of the spontaneously occurring clusters (i.e., $H_{DC} = 0$) under the same coculture conditions. The variability among different experiments (n = 3 for each experiment) remained within the error bars of each individual experiment, showing that the SLP can be effectively increased *in vitro* by the linear agglomerates formed within the cells under the applied field incubation conditions. The concentration of MNPs in each pellet was similar, supporting our hypothesis that the application of a constant field $H_{DC}$ overnight does not affect the amount of cell uptake. In terms of the 'macroscopic' SLP measured in the pellets, the averages of the complete set of experiments (n = 9) in each condition gave SLP=305±25 W/g for H = 0 (clusters) and 576±33 W/g for an applied field of $H_{DC}$ = 650 kA/m (chains), an increase by a factor of ≈2 that is clearly above the experimental error. This value is, to the best of our knowledge, the highest *in vitro* SLP value reported to date. After coculture under an $H_{DC}$ field, the cells were transferred and placed in a ≈200 μL-volume container; therefore, they were partially free to rotate within the liquid culture medium. As a consequence, the linear agglomerates formed during incubation were not perfectly aligned with the applied field $H_{AC}$ during SLP experiments, but since this $H_{AC}$ field has the effect of partially aligning the elongated clusters along its axis, it is expected that during SLP experiments, the actual configuration of the system would be a preferred alignment along the field axis due to the partial rotation of individual cells.

Comparing the increase in SLP values for the clusters in gelatin vs. the parallel-oriented linear agglomerates in BV2 cells, the ratio $R = {SLP_{gel}}/{SLP_{cluster}}$ yielded values of R = 2.1 and R = 1.8 for the gelatin and cells, respectively (see Figures 3 and 6). This somewhat larger SLP increase observed in the gelatin phantoms can be attributed to the better efficiency of chain formation in gels, as can be clearly observed by comparing the TEM images for each case. The predominantly one-dimensional character (i.e., lower fractal dimension) of the chain-like structures formed in gel phantoms, in turn, is probably due to the larger applied field used for the gelatin ($H_{DC}$ = 1.42

MA/m) experiments compared to the *in vitro* constant field of $H_{DC}$ = 0.65 MA/m due to experimental limitations. The lower fractal dimension in gelatin implies a local coordination of the MNPs closer to 2, and therefore, larger SLP values are expected.

Disregarding the difference in $H_{DC}$ values used in the gelatin and *in vitro* experiments, the metabolic pathways within the cytoplasm could be an active factor inhibiting chain formation under an applied magnetic field. To evaluate this hypothesis, we reproduced the previously discussed in-field experiments but cocultured the cells with MNPs at T = 4 ºC. At this temperature, cell metabolism and enzymatic activity diminish, and most chemical reaction rates are lowered, while the viscosity of the cytoplasm increases. The results showed the same increase in SLP values for the in-field-grown cells at 4 ºC with respect to those cocultivated at 37 ºC, within the experimental error (see Figure S11 in the Supporting Information), from SLP = 349±28 W/g for cells grown at H = 0 at 37 ºC (i.e., clusters) to SLP = 566±18 W/g for the linear agglomerates formed at $H_{DC}$≈ 650 kA/m at T = 4 ºC. These results indicate no major effects from metabolic cell activity on the formation of elongated clusters under a field, i.e., the formation of elongated clusters is not hindered (nor favored) by cytoplasmic activity during the uptake process and/or MNP-endosome trafficking, resulting in similar heating performances.

## 4. *Numerical Model*

The formation of linear structures in colloidal MNPs under magnetic fields has been known since the first ferrofluids were made and the new field of physics termed ferrohydrodynamics started [45, 52]. Nevertheless, the complexity of microscopic mechanisms including magnetic dipolar interactions and magneto-rheological effects [53] have so far prevented the development of a detailed model of the expected magnetic relaxation and power absorption in these systems.[54,55] Indeed, linear agglomerates have been reported to form under ac magnetic fields in hyperthermia experiments [44]. The biological interactions under *in vitro* conditions make the modeling of the magnetic relaxation for interacting MNPs even more complex, and therefore some simplifications are required to attempt workable models. The inversion of the magnetic moment of MNPs over the crystalline axis (Néel relaxation) is the dominant mechanism for virtually all *in vitro* SLP experiments since agglomeration and attachment to the cell membrane hinder Brownian rotation. Therefore, the applied ac magnetic field and the dipolar interactions determine the magnetic response of the MNPs at any given temperature. We applied a model based on the classical Stoner–Wohlfarth approach for non-interacting, single-domain particles for determining the equilibrium orientation of the magnetization in the energy landscape, identifying the local's minima and their angular region.[56] The effects of the temperature and the experimental time windows are introduced in our formulation by resolving the master equation with the Arrhenius law. In this way the temperature plays two roles: it assists the magnetic moment to reverse orientation and induces an effective correction in the magnetization originated from thermal fluctuations at the minimum. The time evolution of the populations ($P_0$ and $P_1$) at each minimum (B0 and B1) is given by the following equation:

$$P_0(t + \delta t) = P_0(t) + L[P_0^\infty - P_0(t)], \qquad \text{Eq. 5}$$

where $P_0^\infty$ is the equilibrium value of $P_0(t)$ at $P_0(t=\infty)$, and $L=1-\exp(-\tau_m/\tau)$ is the probability of finding the particle in the SP regime. The parameters $\tau$ and $\tau_m$ are the effective time of magnetic relaxation and measurement time, respectively. Because we treat populations as fractions of particles, they satisfy that $P_0+P_1=1$. With these considerations, the average magnetization $\vec{\mu}$ can be written as:

$$\langle\vec{\mu}\rangle = (1 - L)(P_0 \langle\vec{\mu}\rangle_B^0 + P_1 \langle\vec{\mu}\rangle_B^1) + L \langle\vec{\mu}\rangle_{SP} \qquad \text{Eq. 6}$$

where the brackets represent the thermal statistical average of the magnetization in the superparamagnetic (SP) and blocked (B) regimes. The averages are calculated by integrating over all microstates of the system (for the SP case), at the region that corresponds to each minimum (B0 and B1), or both as in the superparamagnetic case $\langle \vec{\mu} \rangle_{SP}$. Eq. 6 can be interpreted as the average value of the magnetization contributions from the blocked particles (the first term in Eq. 6) where the contributions of both hysteresis minima are considered, and the contribution from superparamagnetic particles (the second term in Eq. 6).

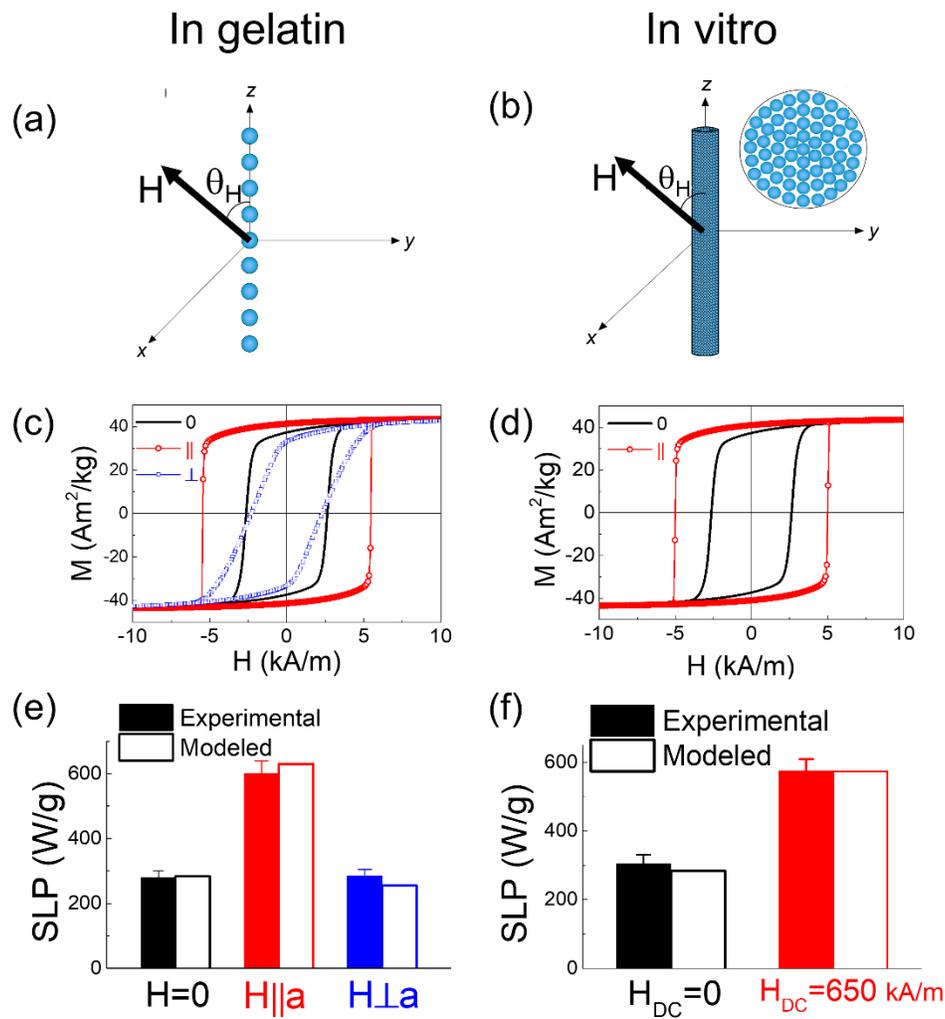

**Figure 7.** Schematic of the geometries used to simulate the SLP of a) linear chains in gelatin and b) lateral view of cylindrical clusters *in vitro* (inset: cross section of the cylindrical agglomerate). The M(H) hysteresis loops calculated under the $H_{AC}$ ac field conditions are also shown for (c) the linear chains in both parallel and perpendicular directions respect to $H_{AC}$ and for (d) the cylindrical clusters (only parallel orientation, see text). The calculated SLP values from these hysteresis loops in (e) gelatin and (f) in vitro for each geometry are shown, as well as the comparison to the experimentally measured SLP.

It is worth noting that for the present low-anisotropy MNPs, the assumption that the anisotropy field $H_K$ is lower than the amplitude of the ac applied field $H_0$ for the experimental conditions is no longer valid. This fact and the interparticle interactions considered make the linear response theory (LRT) no longer applicable. For the model presented here, a more complete analysis on the role of $H_0$ and $f$ parameters as well as applicability ranges have been provided elsewhere.[57]

For the case of interacting systems, the statistical analysis of the clusters was performed assuming a set of chains of identical and equally spaced particles, weighted by the size distribution. In accordance with the microscopy image analysis, three different systems were modeled: a) the 3D clusters 'as obtained' in the gelatin and cells (i.e., without an applied field); b) the 'one-dimensional' chains observed in resins dried under an applied field and c) cylindrical arrangements of MNPs simulating the elongated clusters observed *in vitro* after incubation under a magnetic field. No interactions between different chains or cylinders were considered. **Figure 7** shows the linear and cylindrical geometries used to statistically represent the dipolar interactions actually acting in our systems. For these simulations we considered the anisotropy axes of the clusters to be randomly oriented so that the average magnetic moments of the particles were aligned with the external field. This hypothesis was based on the fact that although the SLP measurements were done on cells with MNPs agglomerates oriented along the applied $H_{DC}$ field, those cells were conditioned and measured immediately after the incubation with culture medium, which allowed the cells to freely rotate.

In the case of MNPs taken up by cells under $H_{DC}= 0$, dipolar interactions between particles are also present. However, in this case, the MNPs are randomly arranged, and the effective dipolar field is statistically reduced. The local dipolar field acting on $i^{th}$ particle was calculated from the following expression:

$$\vec{H}_D^i = \sum_{j \neq i} \left[ \frac{3\vec{r}_{ij}(\langle\vec{\mu}_j\rangle \cdot \vec{r}_{ij})}{r_{ij}^5} - \frac{\langle\vec{\mu}_j\rangle}{r_{ij}^3} \right] \qquad \text{Eq. 7}$$

where $j$ represents neighboring particles, $\vec{r}_{ij}$ is the vector that goes from the center of particle $i$ to the center of particle $j$ and $\langle\vec{\mu}_j\rangle$ is the effective magnetic moment of the $j^{th}$ particle, which is assumed to be oriented along the direction of the local field. Under the above hypotheses, the $j^{th}$ magnetic moment $\mu$ is represented by $\langle\vec{\mu}_j\rangle = \mu\langle\hat{m}\rangle(\sin\theta_H, 0, \cos\theta_H)$, with $\langle\hat{m}\rangle$ being the effective magnetization unit vector. Then, the effective dipolar field acting over the central particle can be expressed as follows:

$$\vec{H}_D = M_s (\delta_r)^3 Q \vec{f} \qquad \text{Eq. 8}$$

where $M_s$ is the saturated magnetization, $\delta_r = \phi/d$ is the ratio between the diameter $\phi$ and the mean distance of the particles, $Q$ is the quality factor and $\vec{f}$ is a vector related to the geometry and orientation of the chain/cylinder with respect to the external magnetic field. (The expressions for $\vec{f}$ and other additional details are given in the Supporting Information.)

The hysteresis loops and SLPs of our MNPs calculated from Eqs. 7 and 8 (**Figure 7**) showed remarkable agreement with the experimental data, which supports our assumption about the influence of the local coordination of the MNPs on magnetic relaxation. In the case of the gelatin and *in vitro* clusters with spherical symmetry (i.e., formed with $H_{DC} = 0$), the MNP anisotropy axes are randomly oriented within a 3D local coordination, and therefore, the resulting effective local dipolar field $H_D$ is smaller than the amplitude of the applied ac field. These conditions are consistent with the observed constant SLP values measured in gelatin samples as a function of MNP concentration (not shown), suggesting that the decrease in SLP values from the *as-prepared* colloids to the immobilized MNPs in gelatin should be mostly due to the blocking of Brownian relaxation. The comparison between SLP experimental values and their corresponding numerical values from the hysteresis loops are shown in **Figure 7**e and f. The agreement between simulated and experimental SLP data supports our interpretation of the importance of the agglomeration geometry (through the local coordination of MNPs and dipolar interactions) in determining the SPL values.

Some additional insight regarding the main mechanisms behind the magnetic relaxation in these structures can be obtained from analysis of the free parameters used in our model. The difference in the calculated $\delta_r$ values related to the relative distances in cylindrical clusters ($\delta_r = 1.5$) and 1D chains ($\delta_r = 1.8$) reflects the denser agglomeration in cells than in phantoms observed in TEM images. The values of the quality factor $Q$ obtained from the simulations for 1D chains ($Q = 0.6$) are better than the corresponding values for cylinders ($Q = 0.4$) due to the more marked 1D character of the MNP aggregates in gelatin phantom than *in vitro* elongated clusters. Although the intracellular elongated agglomerates obtained from in-field cell culture have a general uniaxial symmetry, they are far from being cylindrical entities. These results clearly indicate that there is still room for further improvement of the heating power of intracellular MNPs by improving the 'quality' of one-dimensional structures, and future efforts to experimentally improve this value are required. The third parameter computed, i.e., the $\vec{f}$ vector, reflects the statistical relative orientation between the main axis of the chains/cylinders with respect to the applied ac field. The calculated value of the parallel component of $\vec{f}$ in 1D chains was $f_\parallel = 2.517$, somewhat larger than the corresponding $f_\parallel = 2.022$ for cylindrical clusters. The better degree of alignment of 1D chains than cylinders under the same physical conditions also indicates that the former are better heaters. However, the formation of single-stranded chains of MNPs within cells by an applied dc field could be quite difficult to achieve given the spontaneous agglomeration occurring before uptake. Lowering the concentration of MNPs could help to partially avoid this agglomeration, but since the total heat released depends on the total number of nanoparticles, this strategy has obvious limits. In contrast, one-dimensional, rigid structures of MNPs assembled *before* cell uptake could offer clear advantages for *in vitro* heating. It seems clear that the underlying physical mechanisms by which agglomeration controls the final power absorption in MNPs will only be identified through a careful comparison between numerical models and experimental data of the intracellular agglomerates, which indicates the need for more data to complete a comprehensive body of experimental data on both the structure of these clusters and the corresponding SLP performance under actual *in vitro* conditions.

## *5. Conclusions*

The increase of the heating power due to MNP organization into low-dimensional *intracellular* structures, reported in this work, suggests that there is still some room for improving the heating efficiency for clinical applications of Magnetic Fluid Hyperthermia. This improvement must be achieved by proper design of the MNP core to maximize Néel relaxation and by controlling the topological fractal dimension of the aggregates naturally occurring *in vitro* and *in vivo*. The present results clearly showed the enhancement of SLP by the formation of chain-like MNP assemblies *in vitro* and suggested that this strategy can also be applied under *in vivo* conditions. Surely, more experimental and theoretical work is required to unveil the microscopic mechanisms operating under biological conditions to determine which physical and biological parameters should be controlled to obtain optimal aggregate morphologies. We believe that there are two main factors still to be optimized to enhance the thermal response of the MNPs. First, a larger magnetic field $H_{DC}$ could be applied to assess whether this parameter affects the final fractal dimension of the aggregates. This step entails the design of new magnetic applicators compatible with cell viability during incubation (sterile conditions, minimum cell stress containers, etc.). Second, the physicochemical nature of the MNP surface could be modified to favor the free alignment of the magnetic cores to form linear chains, increasing the SLP of the system. In this regard, it could be interesting to test the in-field uptake of previously fabricated linear chains by cells, which should improve the thermal outcome for specific external ac field directions. In any case, the development of a comprehensive theoretical model will be a necessary step before developing a physical design of MNPs that can expand their heating efficiency to the next level in clinical magnetic hyperthermia therapy.

ASSOCIATED CONTENT

**Supporting Information**. Supporting Information file is available.

The following files are available free of charge.

MovieH=0.mpg (mpg file type); Movie1.mpg (mpg file type); Movie2.mpg (mpg file type)

AUTHOR INFORMATION

**Corresponding Author**

\* Gerardo F. Goya. University of Zaragoza, Spain. Email: goya@unizar.es

**Present Addresses**

◊ nB Nanoscale Biomagnetics S.L., Zaragoza, Spain.

א Departamento de Física, Universidade Federal de Santa Catarina, Florianópolis, SC, Brazil.

**Author Contributions**

‡ These authors contributed equally. The manuscript was written and improved through contributions of all authors. All authors have given approval to the final version of the manuscript.

**Funding Sources**

The Spanish Ministerio de Ciencia, Innovación y Universidades (Projects RTC-2017-6620-1 and MAT2016-78201-P). The Aragon Regional Government (DGA, Project No. E28_20R). The Argentinian governmental agency ANPCyT (Projects PICT-2016-0288 and PICT-2015-0883). The H2020-MSCA-RISE-2016 project SPICOLOST. CNPq (Grant No. PDE 202340/2015-5).

**Notes**

The raw data required to reproduce these findings are available to download from https://data.mendeley.com/datasets/vp9f3yrc9k/draft?preview=1 . The processed data required to reproduce these findings are available to download from https://data.mendeley.com/datasets/vp9f3yrc9k/draft?preview=1 (DOI: 10.17632/vp9f3yrc9k.2).

ACKNOWLEDGMENTS

The authors are indebted to Dr. R. Fernández-Pacheco and Dr. A. Ibarra from the Laboratorio de Microscopias Avanzadas (LMA) for their advice and technical proficiency with EDX-HRTEM analysis. R. Cabreira Gomes is indebted to CNPq for a postdoctoral fellowship.

ABBREVIATIONS

SLP, Specific Loss Power; MNPs, Magnetic Nanoparticles; PEI, Polyethylenimine; DMEM, Dulbecco's modified Eagle's medium; EDX, Energy dispersive X-ray; SAED-HRTEM, Selected

Area Electron Diffraction-High-Resolution TEM; ZFC, Zero-Field Cooling; FC, Field Cooling; FMR, Ferromagnetic Resonance.

REFERENCES

1. Ito, A.; Honda, H.; Kobayashi, T., Cancer immunotherapy based on intracellular hyperthermia using magnetite nanoparticles: a novel concept of "heat-controlled necrosis" with heat shock protein expression. *Cancer Immunol. Immunother.* **2006,** *55* (3), 320–8.

2. Maier-Hauff, K.; Ulrich, F.; Nestler, D.; Niehoff, H.; Wust, P.; Thiesen, B.; Orawa, H.; Budach, V.; Jordan, A., Efficacy and safety of intratumoral thermotherapy using magnetic iron-oxide nanoparticles combined with external beam radiotherapy on patients with recurrent glioblastoma multiforme. *J. Neuro-Oncol.* **2011,** *103* (2), 317–24.

3. Thiesen, B.; Jordan, A., Clinical applications of magnetic nanoparticles for hyperthermia. *Int. J. Hyperthermia* **2008,** *24* (6), 467–74.

4. Johannsen, M.; Gneveckow, U.; Eckelt, L.; Feussner, A.; WaldÖFner, N.; Scholz, R.; Deger, S.; Wust, P.; Loening, S. A.; Jordan, A., Clinical hyperthermia of prostate cancer using magnetic nanoparticles: presentation of a new interstitial technique. *Int. J. Hyperthermia* **2009,** *21* (7), 637–647.

5. Bordet, A.; Lacroix, L.-M.; Fazzini, P.-F.; Carrey, J.; Soulantica, K.; Chaudret, B., Magnetically induced continuous $CO_2$ hydrogenation using composite iron carbide nanoparticles of exceptionally high heating power. *Angew. Chem. Int. Ed. Engl.* **2016,** *128* (51), 16126–16130.

6. Sanz, B.; Calatayud, M. P.; de Biasi, E.; Lima, E., Jr.; Mansilla, M. V.; Zysler, R. D.; Ibarra, M. R.; Goya, G. F., *In silico* before *in vivo*: how to predict the heating efficiency of magnetic nanoparticles within the intracellular space. *Sci. Rep.* **2016,** *6*, 38733.

7. Martinez-Boubeta, C.; Simeonidis, K.; Makridis, A.; Angelakeris, M.; Iglesias, O.; Guardia, P.; Cabot, A.; Yedra, L.; Estrade, S.; Peiro, F.; Saghi, Z.; Midgley, P. A.; Conde-Leboran, I.; Serantes, D.; Baldomir, D., Learning from nature to improve the heat generation of iron-oxide nanoparticles for magnetic hyperthermia applications. *Sci. Rep.* **2013,** *3*, 1652.

8. Guardia, P.; di Corato, R.; Lartigue, L.; Wilhelm, C.; Espinosa, A.; Garcia-Hernandez, M.; Gazeau, F.; Manna, L.; Pellegrino, T., Water-soluble iron oxide nanocubes with high values of specific absorption rate for cancer cell hyperthermia treatment. *ACS Nano* **2012,** *6* (4), 3080–91.

9. di Corato, R.; Espinosa, A.; Lartigue, L.; Tharaud, M.; Chat, S.; Pellegrino, T.; Menager, C.; Gazeau, F.; Wilhelm, C., Magnetic hyperthermia efficiency in the cellular environment for different nanoparticle designs. *Biomaterials* **2014,** *35* (24), 6400–11.

10. Torres, T. E.; Lima, E.; Calatayud, M. P.; Sanz, B.; Ibarra, A.; Fernandez-Pacheco, R.; Mayoral, A.; Marquina, C.; Ibarra, M. R.; Goya, G. F., The relevance of brownian relaxation as power absorption mechanism in magnetic hyperthermia. *Sci. Rep.* **2019,** *9* (1), 3992.

11. Cabrera, D.; Lak, A.; Yoshida, T.; Materia, M. E.; Ortega, D.; Ludwig, F.; Guardia, P.; Sathya, A.; Pellegrino, T.; Teran, F. J., Unraveling viscosity effects on the hysteresis losses of magnetic nanocubes. *Nanoscale* **2017,** *9* (16), 5094–5101.

12. Zubarev, A. Y., Magnetic hyperthermia in a system of immobilized magnetically interacting particles. *Phys. Rev. E* **2019,** *99* (6-1), 062609.

13. Serantes, D.; Baldomir, D.; Martinez-Boubeta, C.; Simeonidis, K.; Angelakeris, M.; Natividad, E.; Castro, M.; Mediano, A.; Chen, D. X.; Sanchez, A.; Balcells, L. I.; Martínez, B., Influence of dipolar interactions on hyperthermia properties of ferromagnetic particles. *J. Appl. Phys.* **2010,** *108* (7), 073918.

14. Mehdaoui, B.; Tan, R. P.; Meffre, A.; Carrey, J.; Lachaize, S.; Chaudret, B.; Respaud, M., Increase of magnetic hyperthermia efficiency due to dipolar interactions in low-anisotropy magnetic nanoparticles: theoretical and experimental results. *Phys. Rev. B* **2013,** *87* (17), 174419.

15. Fu, R.; Yan, Y.; Roberts, C.; Liu, Z.; Chen, Y., The role of dipole interactions in hyperthermia heating colloidal clusters of densely-packed superparamagnetic nanoparticles. *Sci. Rep.* **2018,** *8* (1), 4704.

16. Zubarev, A. Y.; Iskakova, L. Y.; Safronov, A. P.; Krekhno, R. V.; Kuznetsov, D. K.; Beketov, I. V., Positive feedback of interparticle interaction on magnetic hyperthermia. *J. Magn. Magn. Mater.* **2019,** *489* (1), 165402.

17. Ralandinliu Kahmei, R. D.; Borah, J. P., Clustering of MnFe2O4 nanoparticles and the effect of field intensity in the generation of heat for hyperthermia application. *Nanotechnology* **2018,** *30* (3), 035706.

18. Hovorka, O., Thermal activation in statistical clusters of magnetic nanoparticles. *J. Phys. D-Appl. Phys.* **2017,** *50* (4), 044004.

19. Niculaes, D.; Lak, A.; Anyfantis, G. C.; Marras, S.; Laslett, O.; Avugadda, S. K.; Cassani, M.; Serantes, D.; Hovorka, O.; Chantrell, R.; Pellegrino, T., Asymmetric assembling of iron oxide nanocubes for improving magnetic hyperthermia performance. *ACS Nano* **2017,** *11* (12), 12121–12133.

20. Campanini, M.; Ciprian, R.; Bedogni, E.; Mega, A.; Chiesi, V.; Casoli, F.; Fernandez, C. J.; Rotunno, E.; Rossi, F.; Secchi, A.; Bigi, F.; Salviati, G.; Magen, C.; Grillo, V.; Albertini, F., Lorentz microscopy sheds light on the role of dipolar interactions in magnetic hyperthermia. *Nanoscale* **2015,** *7* (17), 7717–25.

21. Zubarev, A. Y., Effect of internal chain-like structures on magnetic hyperthermia in non-liquid media. *Philos. Trans. R. Soc. A* **2019,** *377* (2143), 20180213.

22. Zubarev, A. Y., Magnetic hyperthermia in a system of ferromagnetic particles, frozen in a carrier medium: effect of interparticle interactions. *Phys. Rev. E* **2018,** *98* (3), 032610.

23. Albarqi, H. A.; Wong, L. H.; Schumann, C.; Sabei, F. Y.; Korzun, T.; Li, X.; Hansen, M. N.; Dhagat, P.; Moses, A. S.; Taratula, O.; Taratula, O., Biocompatible nanoclusters with high heating efficiency for systemically delivered magnetic hyperthermia. *ACS Nano* **2019,** *13* (6), 6383–6395.

24. Avugadda, S. K.; Materia, M. E.; Nigmatullin, R.; Cabrera, D.; Marotta, R.; Cabada, T. F.; Marcello, E.; Nitti, S.; Artés-Ibañez, E. J.; Basnett, P.; Wilhelm, C.; Teran, F. J.; Roy, I.; Pellegrino, T., Esterase-cleavable 2D assemblies of magnetic iron oxide nanocubes: exploiting enzymatic polymer disassembling to improve magnetic hyperthermia heat losses. *Chem. Mater.* **2019,** *31* (15), 5450–5463.

25. Southern, P.; Pankhurst, Q. A., Commentary on the clinical and preclinical dosage limits of interstitially administered magnetic fluids for therapeutic hyperthermia based on current practice and efficacy models. *Int. J. Hyperthermia* **2018,** *34* (6), 671–686.

26. Massart, R.; Cabuil, V., Synthèse en milieu alcalin de magnétite colloïdale: contrôle du rendement et de la taille des particules. *J. Chim. Phys.* **2017,** *84*, 967–973.

27. Vergés, M. A.; Costo, R.; Roca, A. G.; Marco, J. F.; Goya, G. F.; Serna, C. J.; Morales, M. P., Uniform and water stable magnetite nanoparticles with diameters around the monodomain–multidomain limit. *J. Phys. D-Appl. Phys.* **2008,** *41* (13), 134003.

28. Calatayud, M. P.; Soler, E.; Torres, T. E.; Campos-Gonzalez, E.; Junquera, C.; Ibarra, M. R.; Goya, G. F., Cell damage produced by magnetic fluid hyperthermia on microglial BV2 cells. *Sci. Rep.* **2017,** *7* (1), 8627.

29. Sanz, B.; Calatayud, M. P.; Torres, T. E.; Fanarraga, M. L.; Ibarra, M. R.; Goya, G. F., Magnetic hyperthermia enhances cell toxicity with respect to exogenous heating. *Biomaterials* **2017,** *114*, 62–70.

30. Fisker, R.; Carstensen, J. M.; Hansen, M. F.; Bødker, F.; Mørup, S., Estimation of nanoparticle size distributions by image analysis. *Journal of Nanoparticle Research* **2000,** *2* (3), 267-277.

31. Torres, T. E.; Lima, E.; Mayoral, A.; Ibarra, A.; Marquina, C.; Ibarra, M. R.; Goya, G. F., Validity of the Néel-Arrhenius model for highly anisotropic $Co_xFe_{3-x}O_4$ nanoparticles. *J. Appl. Phys.* **2015,** *118* (18), 183902.

32. Giannaccini, M.; Calatayud, M. P.; Poggetti, A.; Corbianco, S.; Novelli, M.; Paoli, M.; Battistini, P.; Castagna, M.; Dente, L.; Parchi, P.; Lisanti, M.; Cavallini, G.; Junquera, C.; Goya, G. F.; Raffa, V., Magnetic nanoparticles for efficient delivery of growth factors: stimulation of peripheral nerve regeneration. *Advanced healthcare materials* **2017,** *6* (7), 1601429.

33. Calatayud, M. P.; Sanz, B.; Raffa, V.; Riggio, C.; Ibarra, M. R.; Goya, G. F., The effect of surface charge of functionalized $Fe_3O_4$ nanoparticles on protein adsorption and cell uptake. *Biomaterials* **2014,** *35* (24), 6389–99.

34. Riggio, C.; Calatayud, M. P.; Hoskins, C.; Pinkernelle, J.; Sanz, B.; Torres, T. E.; Ibarra, M. R.; Wang, L.; Keilhoff, G.; Goya, G. F.; Raffa, V.; Cuschieri, A., Poly-l-lysine-coated magnetic nanoparticles as intracellular actuators for neural guidance. *International journal of nanomedicine* **2012,** *7*, 3155–66.

35. Shenker, H., Magnetic anisotropy of cobalt ferrite ($Co_{1.01}Fe_{2.00}O_{3.62}$) and nickel cobalt ferrite ($Ni_{0.72}Fe_{0.20}Co_{0.08}Fe_2O_4$). *Phys. Rev.* **1957,** *107* (5), 1246–1249.

36. Lacroix, L. M.; Lachaize, S.; Hue, F.; Gatel, C.; Blon, T.; Tan, R. P.; Carrey, J.; Warot-Fonrose, B.; Chaudret, B., Stabilizing vortices in interacting nano-objects: a chemical approach. *Nano Lett.* **2012,** *12* (6), 3245–50.

37. Bautin, V. A.; Seferyan, A. G.; Nesmeyanov, M. S.; Usov, N. A., Properties of polycrystalline nanoparticles with uniaxial and cubic types of magnetic anisotropy of individual grains. *Journal of Magnetism and Magnetic Materials* **2018,** *460*, 278-284.

38. Usov, N.; Nesmeyanov, M.; Tarasov, V., Magnetic vortices as efficient nano heaters in magnetic nanoparticle hyperthermia. *Scientific reports* **2018,** *8* (1), 1-9.

39. Usov, N. A.; Nesmeyanov, M. S.; Gubanova, E. M.; Epshtein, N. B., Heating ability of magnetic nanoparticles with cubic and combined anisotropy. *Beilstein J. Nanotechnol.* **2019,** *10*, 305–314.

40. Aquino, V. R. R.; Figueiredo, L. C.; Coaquira, J. A. H.; Sousa, M. H.; Bakuzis, A. F., Magnetic interaction and anisotropy axes arrangement in nanoparticle aggregates can enhance or reduce the effective magnetic anisotropy. *Journal of Magnetism and Magnetic Materials* **2020,** *498*, 166170.

41. Liu, X. L.; Choo, E. S. G.; Ahmed, A. S.; Zhao, L. Y.; Yang, Y.; Ramanujan, R. V.; Xue, J. M.; Fan, H. M.; Ding, J., Magnetic nanoparticle-loaded polymer nanospheres as magnetic hyperthermia agents. *Journal of Materials Chemistry B* **2014,** *2* (1), 120-128.

42. Hayes, C. F.; Hwang, S. R., Observation of magnetically induced polarization in a ferrofluid. *J. Colloid Interface Sci.* **1977,** *60* (3), 443–447.

43. Martinet, A., Birfringence et dichrosme linaire des ferrofluides sous champ magntique. *Rheol. Acta* **1974,** *13* (2), 260–264.

44. Saville, S. L.; Qi, B.; Baker, J.; Stone, R.; Camley, R. E.; Livesey, K. L.; Ye, L.; Crawford, T. M.; Mefford, O. T., The formation of linear aggregates in magnetic hyperthermia: implications on specific absorption rate and magnetic anisotropy. *J. Colloid Interface Sci.* **2014,** *424*, 141–51.

45. Hayes, C. F., Observation of association in a ferromagnetic colloid. *J. Colloid Interface Sci.* **1975,** *52* (2), 239–243.

46. Balcells, L.; Stankovic, I.; Konstantinovic, Z.; Alagh, A.; Fuentes, V.; Lopez-Mir, L.; Oro, J.; Mestres, N.; Garcia, C.; Pomar, A.; Martinez, B., Spontaneous in-flight assembly of magnetic nanoparticles into macroscopic chains. *Nanoscale* **2019,** *11* (30), 14194–14202.

47. Darras, A.; Opsomer, E.; Vandewalle, N.; Lumay, G., Superparamagnetic colloids in viscous fluids. *Sci. Rep.* **2017,** *7* (1), 7778.

48. Beck, M. M.; Lammel, C.; Gleich, B., Improving heat generation of magnetic nanoparticles by pre-orientation of particles in a static three tesla magnetic field. *J. Magn. Magn. Mater.* **2017,** *427*, 195–199.

49. Jeon, S.; Hurley, K. R.; Bischof, J. C.; Haynes, C. L.; Hogan, C. J., Quantifying intra-and extracellular aggregation of iron oxide nanoparticles and its influence on specific absorption rate. *Nanoscale* **2016,** *8* (35), 16053–64.

50. Liu, X.; Zheng, J.; Sun, W.; Zhao, X.; Li, Y.; Gong, N.; Wang, Y.; Ma, X.; Zhang, T.; Zhao, L. Y.; Hou, Y.; Wu, Z.; Du, Y.; Fan, H.; Tian, J.; Liang, X. J., Ferrimagnetic vortex nanoring-mediated mild magnetic hyperthermia imparts potent immunological effect for treating cancer metastasis. *ACS Nano* **2019,** *13* (8), 8811–8825.

51. Serantes, D.; Simeonidis, K.; Angelakeris, M.; Chubykalo-Fesenko, O.; Marciello, M.; Morales, M. D. P.; Baldomir, D.; Martinez-Boubeta, C., Multiplying magnetic hyperthermia response by nanoparticle assembling. *The Journal of Physical Chemistry C* **2014,** *118* (11), 5927-5934.

52. Neuringer, J. L.; Rosensweig, R. E., Ferrohydrodynamics. *Phys. Fluids* **1964,** *7* (12), 1927–1937.

53. Furst, E. M.; Gast, A. P., Dynamics and lateral interactions of dipolar chains. *Phys. Rev. E Stat. Phys. Plasmas Fluids Relat. Interdiscip. Top.* **2000,** *62* (5 Pt B), 6916–25.

54. Iglesias, O., Magnetic nanoparticle assemblies from fabrication to clinical applications. In *Theory to Therapy Chemistry to Clinic Bench to Bedside*, Trohidou, K. N., Ed. CRC Press: Boca Raton, FL, 2015; pp 301–306.

55. Vernay, F.; Sabsabi, Z.; Kachkachi, H., Ac susceptibility of an assembly of nanomagnets: combined effects of surface anisotropy and dipolar interactions. *Phys. Rev. B* **2014,** *90* (9), 094416.

56. de Biasi, E.; Zysler, R. D.; Ramos, C. A.; Knobel, M., A new model to describe the crossover from superparamagnetic to blocked magnetic nanoparticles. *J. Magn. Magn. Mater.* **2008,** *320* (14), e312–e315.

57. Valdés, D. P.; Lima, E.; Zysler, R. D.; De Biasi, E., Modeling the Magnetic-Hyperthermia Response of Linear Chains of Nanoparticles with Low Anisotropy: A Key to Improving Specific Power Absorption. *Physical Review Applied* **2020,** *14* (1), 014023.

**Supporting Information**

Low Dimensional Assemblies of Magnetic MnFe$_2$O$_4$ Nanoparticles and Direct *In Vitro* Measurements of Enhanced Heating Driven by Dipolar Interactions: Implications for Magnetic Hyperthermia

*Beatriz Sanz*[†,§,◊,‡,1]*, Rafael Cabreira-Gomes*[†,ℵ,‡,2]*, Teobaldo E. Torres*[†,¶,⁎,3]*, Daniela P. Valdés*[⁎,4]*, Enio Lima Jr.*[⁎,5]*, Emilio De Biasi*[⁎,6]*, Roberto D. Zysler*[⁎,7]*, M. Ricardo Ibarra*[†,§,8] *and Gerardo F. Goya*[†,§,⁎,9]*.*

[†] Instituto de Nanociencia de Aragón, Universidad de Zaragoza, Zaragoza, Spain

[§] Condensed Matter Physics Department, University of Zaragoza, Zaragoza, Spain

[¶] Laboratorio de Microscopias Avanzadas, Universidad de Zaragoza, Zaragoza, Spain

[⁎] Comisión Nacional de Investigaciones Científicas y Técnicas (CONICET), Centro Atómico Bariloche, Bariloche, Argentina.

[1] Email: bsanzsague@gmail.com

[2] Email: rafaelfsc@gmail.com

[3] Email: teobaldotorresmolina@gmail.com

[4] Email: valdes.danip@gmail.com

[5] Email: eniolimajr@gmail.com

[6] Email: debiasiem@gmail.com

[7] Email: zysler@cab.cnea.gov.ar

[8] Email: ibarra@unizar.es

[9] Email: goya@unizar.es

**Table S1.** Parameters of MnFe$_2$O$_4$ MNPs from Three Different Batches. The average particle Diameter <d> and standard deviation σ were obtained from Gaussian fits of each of the particle distributions. The atomic ratios ρ=[Fe]/[Mn] were obtained from EDS-SEM and EDS-STEM analyses, Yielding the nominal atomic compositions, Mn$_x$Fe$_{3-x}$O$_4$ is shown in the last column

| Sample | <d> (nm) | σ (nm) | ρ=[Fe]/[Mn] EDS-SEM | ρ=[Fe]/[Mn] EDS-STEM | Mn$_x$Fe$_{3-x}$O$_4$ |
|---|---|---|---|---|---|
| MnFe$_2$O$_4$ | 46 | 15 | 2.14±6 | 2.04±8 | Mn$_{0.97}$Fe$_{2.03}$O$_4$ |
| MnFe$_2$O$_4$ | 49 | 11 | 2.03±4 | 1.99±6 | Mn$_{0.99}$Fe$_{2.00}$O$_4$ |
| MnFe$_2$O$_4$ | 51 | 13 | 2.11±6 | 2.07±8 | Mn$_{0.97}$Fe$_{2.03}$O$_4$ |

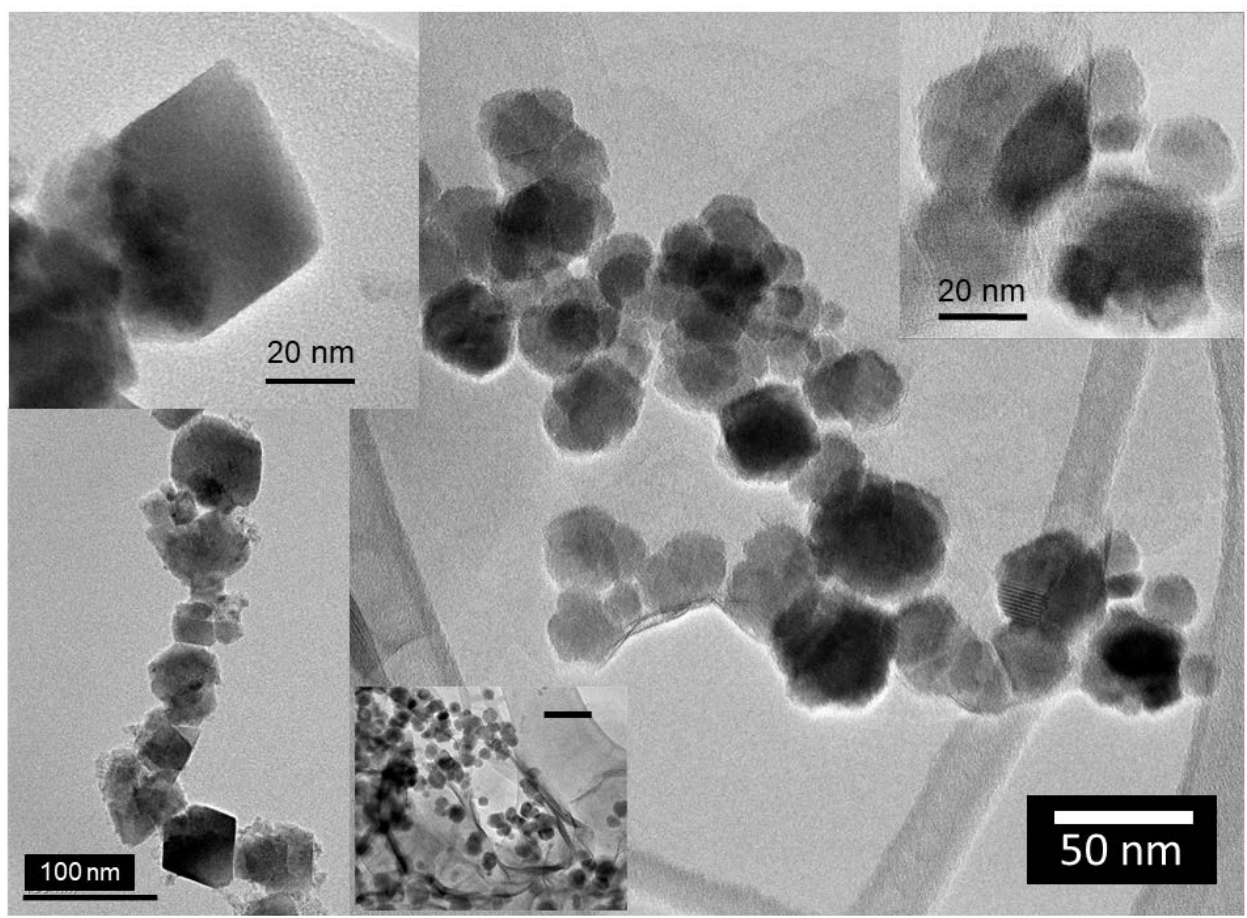

**Figure S1.** Bright-field transmission electron microscopy images of MnFe$_2$O$_4$ MNPs from different batches.

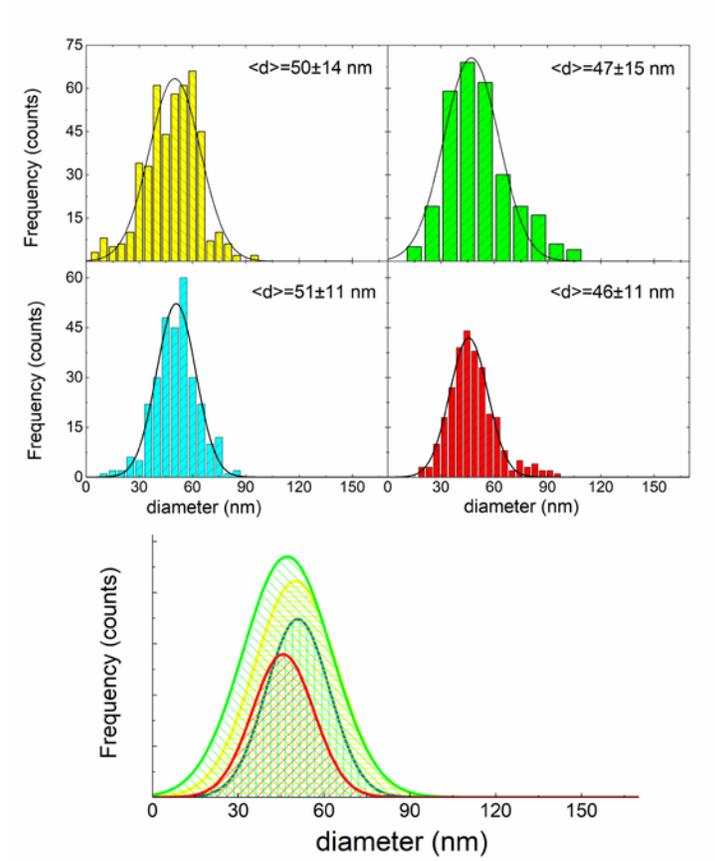

**Figure S2.** Histograms from TEM images of four different samples. The solid lines are the best fits using a Gaussian distribution. Each of the four panels shows the obtained mean values and standard deviations. The bottom graph shows superposition of the four Gaussian distributions for comparative purposes.

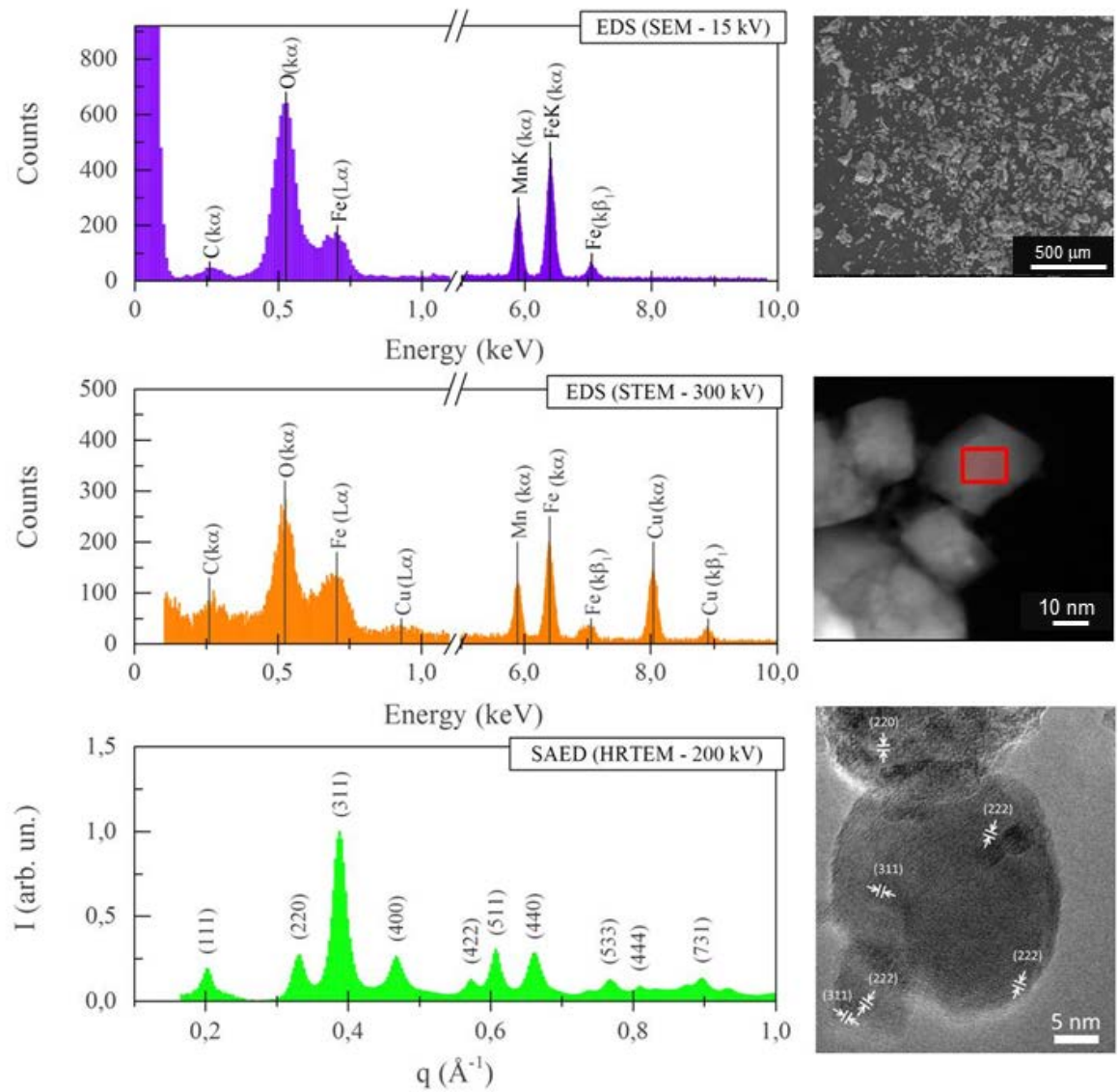

**Figure S3.** EDS analysis of MnFe$_2$O$_4$ MNPs.

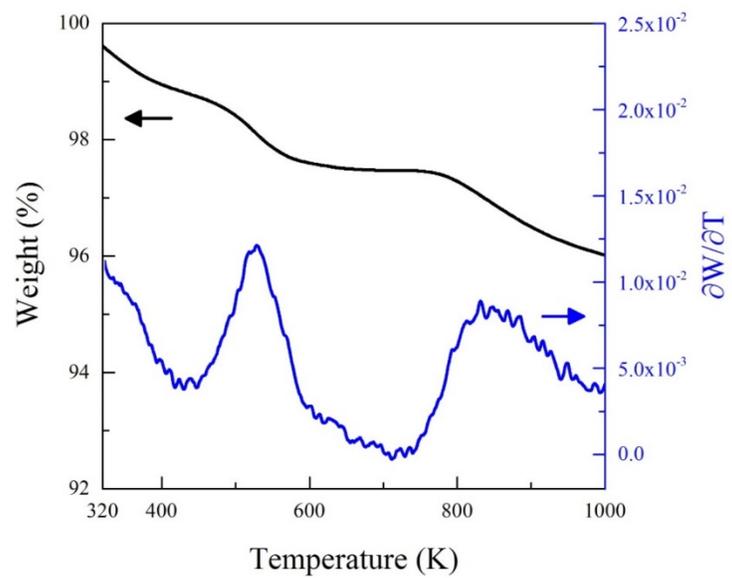

**Figure S4.** Thermogravimetric curve of MnFe$_2$O$_4$ nanoparticles.

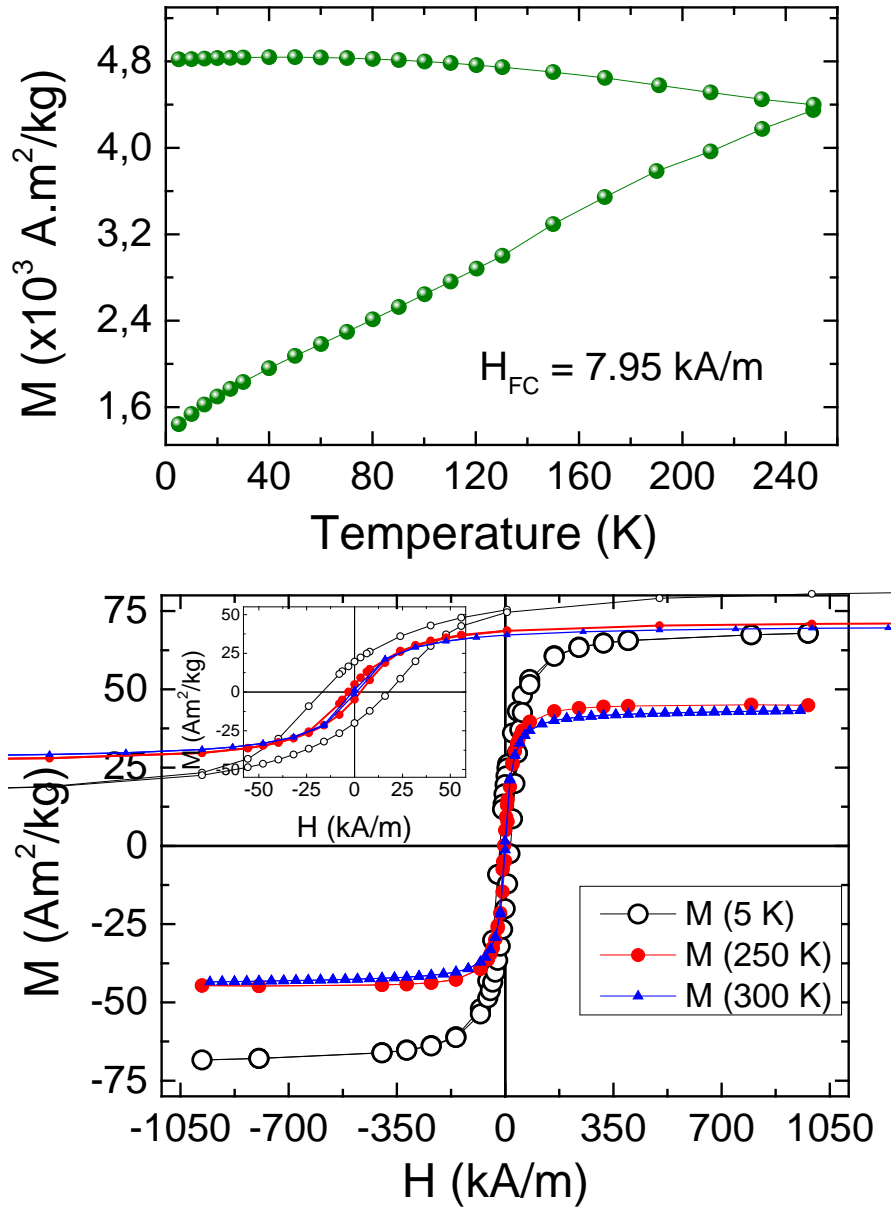

**Figure S5.** Magnetization data for MnFe$_2$O$_4$ nanoparticles. Upper panel: magnetization *vs.* temperature data, taken in ZFC and FC modes, with H$_{FC}$ = 7.95 kA/m (100 Oe). The data were collected with increasing temperature in both curves. Lower panel: hysteresis loops of M(H) at different temperatures. The inset shows the low-field region, enlarged to display the different coercive fields H$_C$ at each temperature.

**FITTING PROCEDURE FOR H$_C$(T)**

The thermally activated model for magnetic relaxation in single-domain magnetic nanoparticles is given by $H_C(T) = \frac{2K_{eff}}{M_S}\left[1 - \left(\frac{T}{T_B}\right)^{1/2}\right]$.[1,2] When the blocking temperatures of the MNPs are close to or above room temperature, H$_C$(T) will be nonzero for a wide range of temperatures (in our case, from 5 to 300 K). Since K$_{eff}$ can change by up to one order of magnitude within this temperature range, the H$_C$(T) expression above must include the thermal dependence of this parameter. Incorporating this dependence into the H$_C$(T) expression and considering $K_{eff}$(T) as the first magnetocrystalline anisotropy constant $K_1$(T), the coercive field of a particle of volume V and saturation magnetization M$_S$ is given by the following:

$$H_C(T) = 0.48 \frac{2K_1(0)\, e^{-BT^2}}{\mu_0 M_S}\left[1 - \left(\frac{25 k_B T}{V A e^{-BT^2}}\right)^{1/2}\right] \quad \text{Eq. S1}$$

where A and B are fitting parameters, and k$_B$ is the Boltzmann constant. This expression provided a good fit of the experimental data for a limited temperature range (150 K – 300 K, see Figure S6) below room temperature, i.e., in the temperature window that includes the point where the measurements of magnetic hyperthermia and Ferromagnetic Resonance (FMR) were performed. The same expression was unable to fit the experimental data at lower temperatures (T < 150 K). A possible explanation can be found in the average size of the MNPs, which is close to the single-domain (SD) to multi-domain (MD) transition, and for these sizes it is known (for example in iron oxide nanoparticles) that the MNPs may have complicated magnetic structures like vortex-like or close-contour magnetic alignment.[3] Therefore, it is expected that close to the SD-MD transition the magnetization reversal does not occur by a coherent mechanism as assumed in the Stoner Wohlfarth model of SD particles.

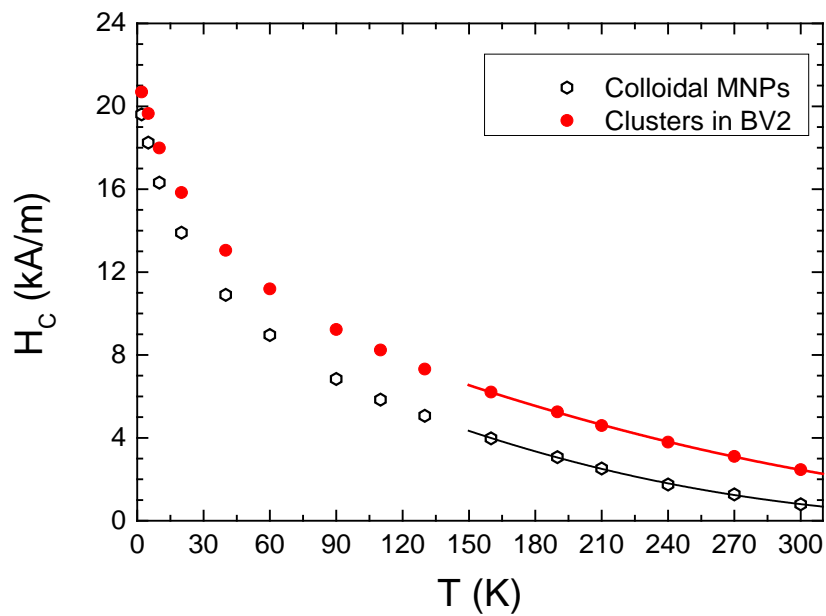

**Figure S6.** Thermal dependence of the coercive field $H_C(T)$ of $MnFe_2O_4$ MNPs in the as-prepared colloids (open black circles) and intracellular clusters within BV2 cells (filled red circles). The solid lines are the fits with the equation S1

**Table S2.** Effective Magnetic Anisotropy $K_{eff}$ Obtained from the Low- and High-Temperature Fitting of the Thermal Dependence of the Coercive Field $H_C(T)$ and from the Angular Dependence of the FMR Resonance Field $H_r(\theta)$

|  | $K_{eff}$ from $H_C(T)$ | | $K_{eff}$ from $H_r$ vs. $\theta$ |
| --- | --- | --- | --- |
|  | T = 2 K | T = 297 K | T = 297 K |
| **Colloidal $MnFe_2O_4$** | 1.6 kJ/m³ | 0.63 kJ/m³ | 3.5 kJ/m³ |
| **Clusters in BV2 cells** | 2.6 kJ/m³ | 1.4 kJ/m³ | -- |

# CRITICAL DIAMETER FOR SINGLE- TO MULTI-DOMAIN TRANSITION

The critical length ($\delta_w$) of magnetic material for formation of a domain wall is given by

$$\delta_w = \pi \sqrt{\frac{A}{K}} \qquad \text{Eq. S2}$$

being $A$ the exchange stiffness and $K$ the effective magnetic anisotropy constant (J/m³). The exchange stiffness (J/m) is given by

$$A = \frac{nES^2}{\langle a \rangle} \qquad \text{Eq. S3}$$

where $n$ is the number of atoms with same spin (in the case of FCC crystalline structure – n = 4), <a> is the lattice parameter, $S$ is the spin and $E$ is the exchange energy, which can be estimated by the Curie temperature ($T_C$), by the relation, $E = 0.3 k_B T_C$.

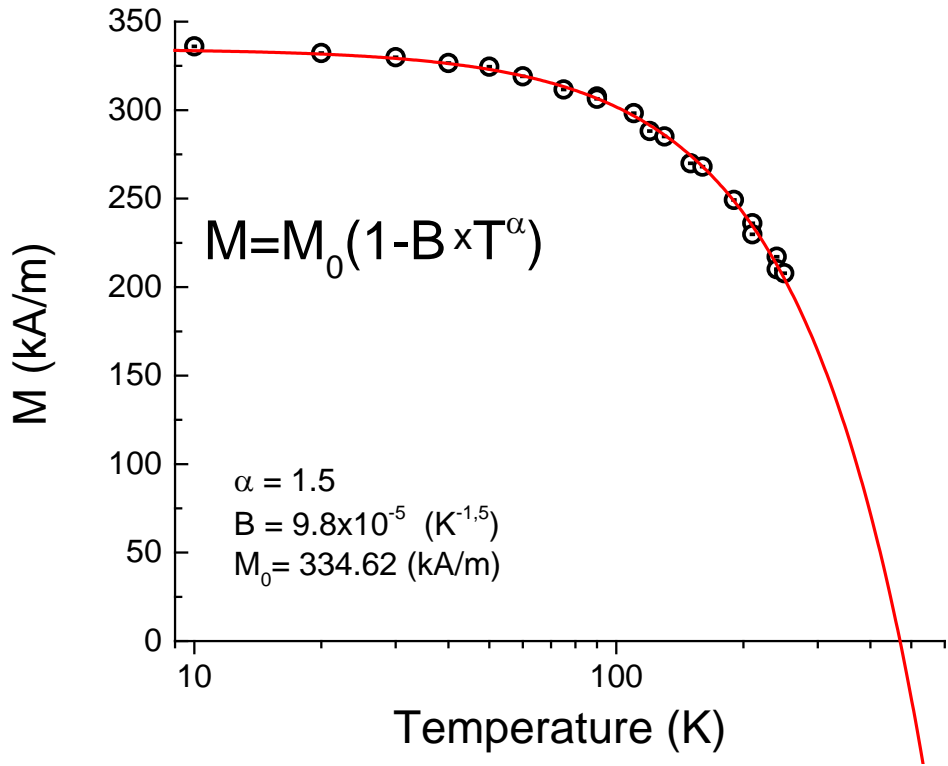

*Figure S7. Thermal dependence of saturation magnetization. The solid line is the best fit using the Bloch's law*

In our case, we have estimated a $T_C \sim 470$ K by using a rough extrapolation of Bloch's law, see fig. S7. Using this value, we obtained $E = 1.94 \times 10^{-21}$ Joules. From the lattice parameter value for MnFe$_2$O$_4$ FCC structure $a = 0.851$ nm and the magnetic anisotropy constant $K_{eff} = 3.5$ kJ/m$^3$ (from FMR data, see below), we calculated an exchange stiffness $\approx 2.29 \times 10^{-12}$ J/m, yielding $\delta_W = 80$ nm.

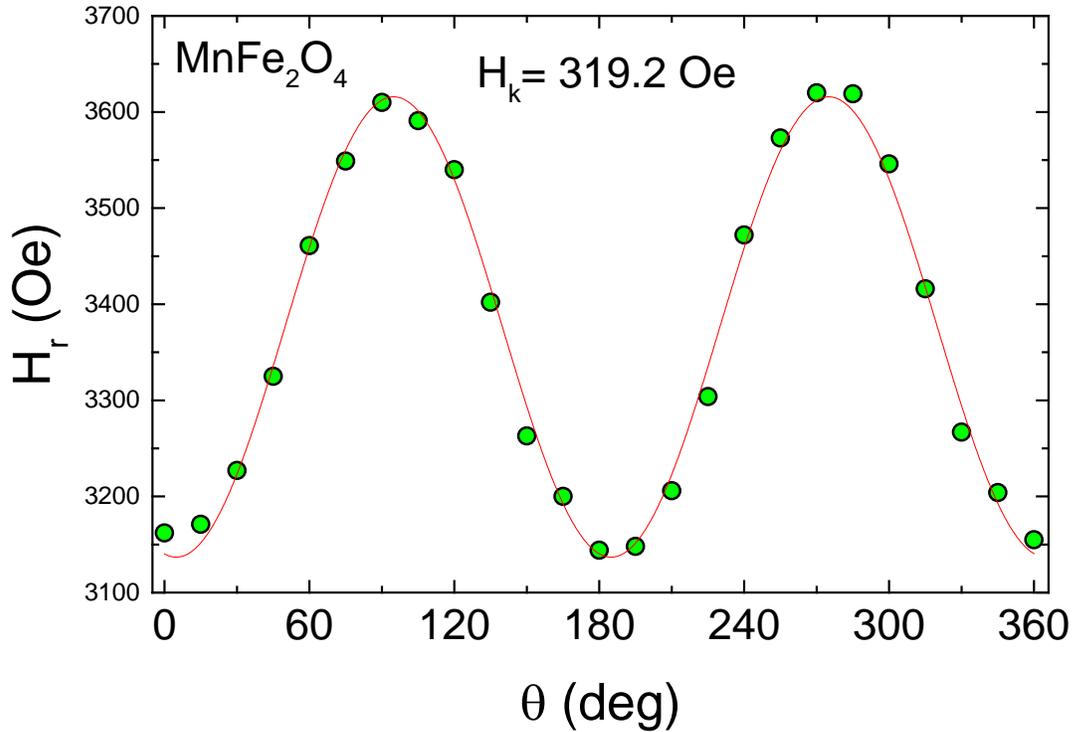

**Figure S8.** Angular dependence of the FMR field of MnFe$_2$O$_4$ MNPs to obtain $K_{eff}$. Samples were conditioned by aligning the anisotropy axis and fixing the MNPs in polyacrylic polymer. The solid line is the fit using the approximation from the Legendre polynomial for the resonance field $H_r = \frac{\omega}{\gamma} - \frac{3}{2} H_A^{eff} [cos^2(\theta) - 1/3]$.

**Power Absorption Experiments.** The SLP values of the *as-prepared* colloids were measured at different magnetic field amplitudes and frequencies ($0 < H_0 < 24$ kA/m; $0 < f < 820$ kHz). The functional dependence of SLP($H_0, f$) on both $H_0$ and $f$ is described by the following expression: [4,5]

$$SLP = \Phi \, H_0^\lambda \frac{\Omega f^2}{(\Omega f)^2 + 1} \qquad \text{Eq. S4}$$

where $\Phi$ is a field- and frequency-independent parameter that includes the magnetic properties of the material, and $\Omega = 2\pi\tau_{eff}$, with $\tau_{eff}$ being the effective relaxation time of the MNPs.

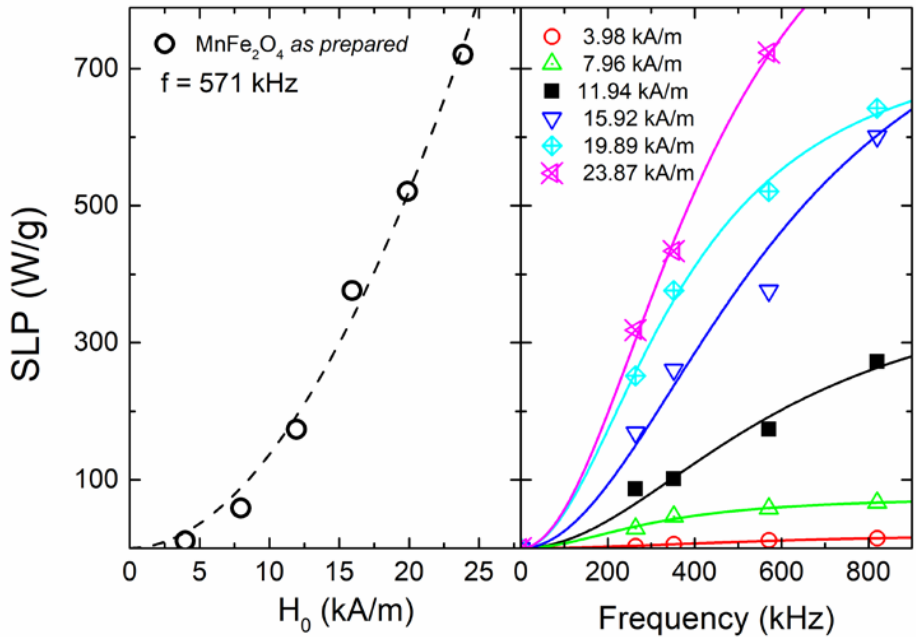

**Figure S9.** SLP values (symbols) of the *as-prepared* colloids. Left panel: samples measured at increasing magnetic field amplitudes of $0 < H_0 < 24$ kA/m (open circles). The dashed line is the best fit using the equation SLP($H_0$) = $AH_0^\lambda$ (see text). Right panel: frequency dependence of SLP at different $H_0$ values (symbols). Solid lines are the best fits using the expression SLP = $\Phi \frac{\Omega f^2}{(\Omega f)^2 + 1}$, as described in the main text.

**Table S3.** Parameters $\Phi$ and $\Omega$ obtained from the fitted frequency dependence of the SLP data in Figure S9, using the equation $\text{SLP} = \Phi \frac{\Omega f^2}{(\Omega f)^2 + 1}$. The parameter $\tau = \frac{\Omega}{2\pi}$ is also given. Data was taken with different field amplitudes of $3.98 \text{ kA/m} \leq H_0 \leq 24 \text{ kA/m}$

| $H_0$ (kA/m) | $\Phi$ | $\Omega$ | $\tau$ (s) |
|---|---|---|---|
| 3.98 | 2.74E-05 | 1.66E-06 | 2.61E-06 |
| 7.96 | 1.77E-04 | 3.21E-06 | 5.04E-06 |
| 11.94 | 4.86E-04 | 1.66E-06 | 2.60E-06 |
| 15.92 | 0.00108 | 1.73E-06 | 2.72E-06 |
| 19.89 | 0.00149 | 2.70E-06 | 4.24E-06 |
| 23.87 | 0.00189 | 2.28E-06 | 3.58E-06 |

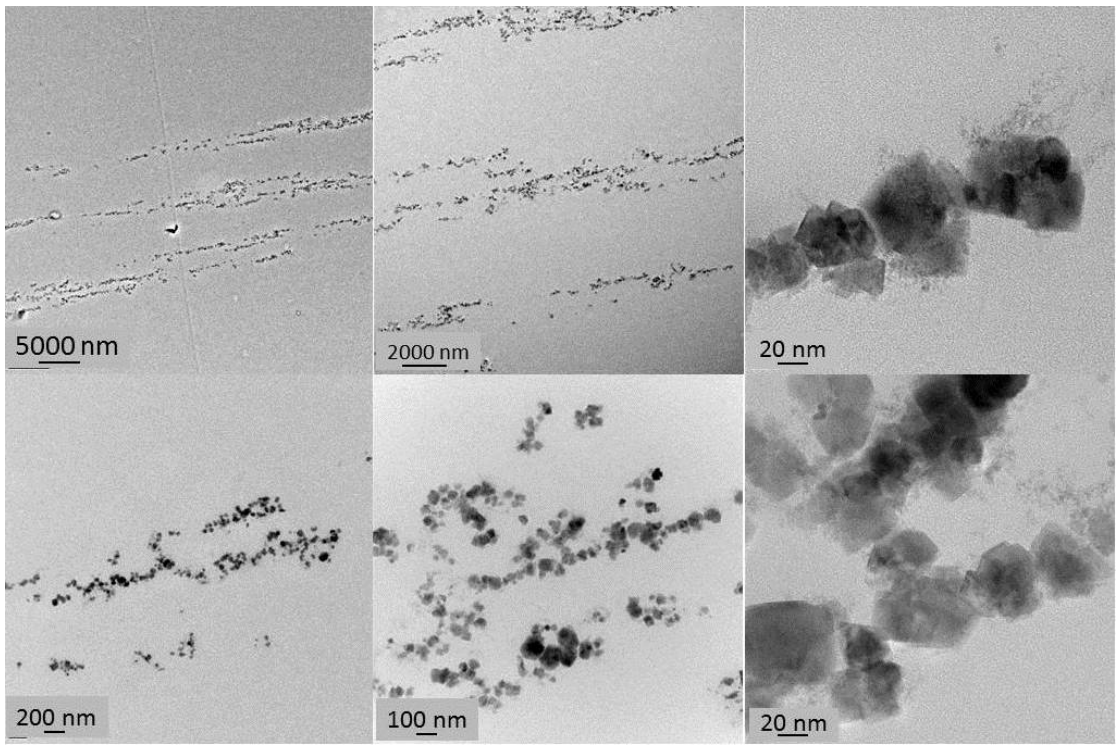

**Figure S10.** TEM images of chain-like clusters of MnFe$_2$O$_4$ MNPs in solid resin grown under a constant dc magnetic field of $H_{DC} = 1.42$ MA/m during solidification.

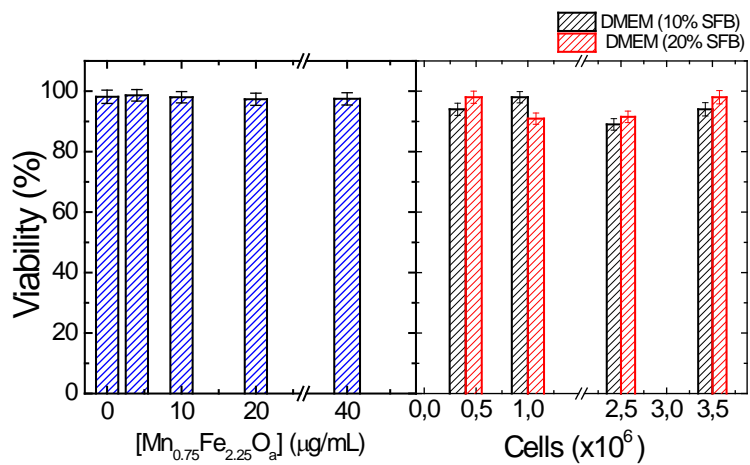

**Figure S11.** Cell viability data of BV2 microglia cells to assess the toxicity of the $MnFe_2O_4$ nanoparticles used in this work. Left panel: viability as a function of MNP concentration. Right panel: two different FBS amounts, 10% (black bars) and 20% (red bars), were used with increasing numbers of BV2 cells to test the conditions for minimum cell stress during incubation with MNPs.

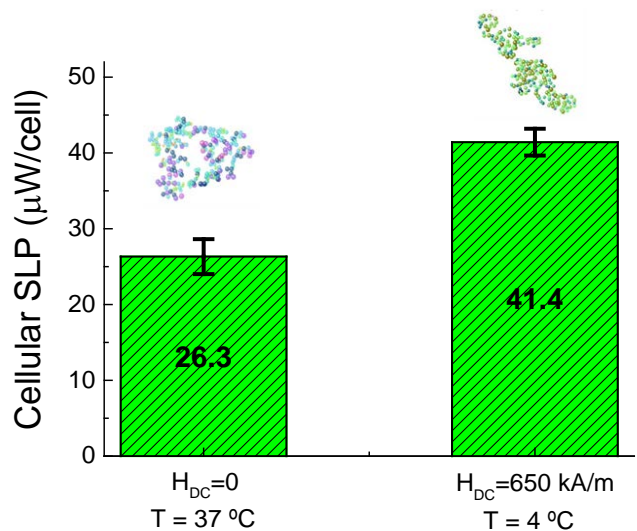

**Figure S12.** SLP of chains formed during culture overnight at 4 °C under an applied field of H = 650 kA/m compared with the SLP of MNP clusters formed in cells cultured at 37 °C and $H_{DC} = 0$ in the same experiment.

**Numerical Calculation of SLP for Aggregates.** The simple model used in this work to simulate the elongated clusters obtained in phantoms and in cells consists of ideal 1D- chains and elongated clusters with cylindrical symmetry, respectively. The local interaction field acting on particle $i$ can be expressed by Equation III given in the main text:

$$\vec{H}_D^i = \sum_{j \neq i} \left[ \frac{3\vec{r}_{ij}(\langle\vec{\mu}_j\rangle \cdot \vec{r}_{ij})}{r_{ij}^5} - \frac{\langle\vec{\mu}_j\rangle}{r_{ij}^3} \right]$$

To simplify the calculations, we assume that the $i^{th}$ particle is located in the geometrical center of each structure (chain or cylinder). Under the hypothesis of our model, the vector $\vec{r}_{ij}$ can be expressed as $\vec{r}_{ij} = d(x_j, y_j, z_j)$, where d is the mean interparticle distance and the coordinates of the vector represent the distance in units of d.

In Eq. (IV), for parallel and perpendicular orientations, $\vec{f}$ has the following expressions:

$$\begin{cases} f_\parallel = \frac{\pi}{6} \left[ \sum_j \frac{3\,z_j^2}{(\rho_j^2+z_j^2)^{5/2}} - \frac{1}{(\rho_j^2+z_j^2)^{3/2}} \right] \\ f_\perp = \frac{\pi}{6} \left[ \sum_j \frac{3\,\rho_j^2}{(\rho_j^2+z_j^2)^{5/2}} - \frac{1}{(\rho_j^2+z_j^2)^{3/2}} \right] \end{cases} \qquad \text{Equation (S5)}$$

where $\rho_j^2 = x_j^2 + y_j^2$. For perfectly linear chains, $\rho_j$ is null.

The parameters for both simulated cases are $M_s$ = 44.4 Am²/kg, K = 1.1·10³ J/m³, a normal size distribution with mean diameter $\langle\phi\rangle = 51$ nm and dispersion $\sigma = 11$ nm. In both cases, we assumed a length of 5 μm for the structures (cylinders and 1D chains). For the cylinders, we assumed a thickness of 500 nm (10 nanoparticles on average). The specific parameters for the simulations are shown in Table S2.

**REFERENCES**

1.	Cullity, B. D., Introduction to Magnetic Materials. In *Addison-Wesley Series in Metallurgy and Materials*, Cullity, B. D., Ed. Addison-Wesley Pub. Co.: Reading, Massachusetts, 1972; p 666.
2.	Torres, T. E.; Lima, E.; Mayoral, A.; Ibarra, A.; Marquina, C.; Ibarra, M. R.; Goya, G. F., Validity of the Néel-Arrhenius Model for Highly Anisotropic Co$_x$Fe$_{3-x}$O$_4$ Nanoparticles. *J. Appl. Phys.* **2015,** *118* (18), 183902.
3.	Bautin, V. A.; Seferyan, A. G.; Nesmeyanov, M. S.; Usov, N. A., Properties of polycrystalline nanoparticles with uniaxial and cubic types of magnetic anisotropy of individual grains. *Journal of Magnetism and Magnetic Materials* **2018,** *460*, 278-284.
4.	Sanz, B.; Calatayud, M. P.; De Biasi, E.; Lima, E., Jr.; Mansilla, M. V.; Zysler, R. D.; Ibarra, M. R.; Goya, G. F., In Silico before *In Vivo*: How to Predict the Heating Efficiency of Magnetic Nanoparticles within the Intracellular Space. *Scientific reports* **2016,** *6*, 38733.
5.	Sanz, B.; Calatayud, M. P.; Torres, T. E.; Fanarraga, M. L.; Ibarra, M. R.; Goya, G. F., Magnetic Hyperthermia Enhances Cell Toxicity with Respect to Exogenous Heating. *Biomaterials* **2017,** *114*, 62-70.